\documentclass[sigconf]{acmart}

\setcopyright{none}
\acmConference[CONF'25]{Conference Name}{Month, 2025}{City, Country}
\acmYear{2025}
\acmBooktitle{CONF '25: Conference Name, Month, 2025, City, Country}
\usepackage{tikz}
\usepackage{amsmath}

\usepackage{amssymb}
\usepackage{filecontents}
\usepackage{multirow}
\usepackage{makecell}
\usepackage{colortbl} %
\usepackage{pifont}
\usepackage{adjustbox}
\usepackage[most]{tcolorbox}
\usepackage{tabularx}
\usepackage{array}
\usepackage{subcaption}
\usepackage{listings}

\newcommand{\bench}{VulnRepairEval}
\newcommand{\codelink}{(Please contact the authors.)}

\begin{document}
	
	\title{\bench: An Exploit-Based Evaluation Framework for Assessing Large Language Model Vulnerability Repair Capabilities}
	
	\author{Weizhe Wang}
	\authornote{Co-first author.}
	\affiliation{%
		\institution{Tianjin University}
		\country{China}
	}
	
	\author{Wei Ma}
	\authornotemark[1]
	\affiliation{%
		\institution{Singapore Management University}
		\country{Singapore}
	}
	
	\author{Qiang Hu}
	\affiliation{%
		\institution{Tianjin University}
		\country{China}
	}
	
	\author{Yao Zhang}
	\authornote{Corresponding author. zzyy@tju.edu.cn, losin@tju.edu.cn}
	\affiliation{%
		\institution{Tianjin University}
		\country{China}
	}
	
	\author{Jianfei Sun}
	\affiliation{%
		\institution{Singapore Management University}
		\country{Singapore}
	}
	
	\author{Bin Wu}
	\affiliation{%
		\institution{Tianjin University}
		\country{China}
	}
	
	\author{Yang Liu}
	\affiliation{%
		\institution{Nanyang Technological University}
		\country{Singapore}
	}
	
	\author{Guangquan Xu}
	\authornotemark[2]
	\affiliation{%
		\institution{Tianjin University}
		\country{China}
	}
	
	\author{Lingxiao Jiang}
	\affiliation{%
		\institution{Singapore Management University}
		\country{Singapore}
	}
	
	\renewcommand{\shortauthors}{Wang and Ma et al.}
	
	\keywords{software vulnerability repair, LLM, exploit-based evaluation, benchmark}
	
	\begin{abstract}
		The adoption of Large Language Models (LLMs) for automated software vulnerability patching has shown promising outcomes on carefully curated evaluation sets. Nevertheless, existing datasets predominantly rely on superficial validation methods rather than exploit-based verification, leading to overestimated performance in security-sensitive applications. This paper introduces \textbf{\bench}, an evaluation framework anchored in functional Proof-of-Concept (PoC) exploits. Our framework delivers a comprehensive, containerized evaluation pipeline that enables reproducible \emph{differential} assessment, where repair success requires the original exploit to fail execution against the modified code. The benchmark construction involved extensive data curation: we processed over $400$ CVEs and approximately 2,500 potential sources to extract a collection of authentic vulnerability instances (23 Python CVEs) amenable to automated testing with working PoCs. Through \bench, we conduct a comprehensive evaluation of 12 popular LLMs and observe a significant performance deficit: even the top-performing model successfully addresses merely 5/23 instances (about 21.7\%), exposing critical weaknesses in security-focused applications. Our failure analysis reveals that most unsuccessful attempts stem from imprecise vulnerability identification and patches containing syntactic or semantic errors. Enhanced prompting strategies and multi-agent approaches yield minimal improvements, with overall effectiveness remaining largely unaffected. This work contributes a stringent, practical evaluation framework for LLM-driven vulnerability remediation and underscores the necessity for assessment protocols that authentically reflect real-world exploitation scenarios.
	\end{abstract}
	
	\maketitle
	
	\section{Introduction}
	Large language models (LLMs) are increasingly used to automate vulnerability repair. 
Without rigorous validation they can create a false sense of safety: a patch may appear to fix the issue while the underlying vulnerability remains exploitable.
This risk is acute in the Python ecosystem, which underpins AI, cloud, and data workloads. 
Passing unit tests or simply running without errors does not establish that a vulnerability is eliminated.
Patches can be plausible yet incorrect; they may pass existing tests but fail to satisfy the true security specification or to generalize beyond narrow inputs.
Real incidents illustrate this danger, including an incomplete PyYAML fix that later enabled code execution (CVE-2020-14343~\cite{cve-2020-14343}, noted as an incomplete fix for CVE-2020-1747), a TIFF decoding overflow in Pillow attributed to an incomplete fix (CVE-2021-25289)~\cite{cve-2021-25289}, and a Django ReDoS linked to an incomplete fix of earlier CVEs (CVE-2024-27351)~\cite{cve-2024-27351}.
This paper asks: 

\emph{\textbf{Do LLM generated patches eliminate the attack surface rather than merely keep the program running?}}

The community has proposed many datasets for evaluating LLM coding ability, ranging from synthetic tasks (e.g., HumanEval, MBPP) to real-bug and project-scale benchmarks (e.g., BugsInPy~\cite{widyasari2020bugsinpy}, Defects4J~\cite{just2014defects4j}, SWE-Bench~\cite{jimenez2024swebench}) and security-oriented corpora.
Most ultimately validate with unit tests or a minimal proof-of-vulnerability (PoV) trigger.
Such checks confirm executability but do not prove that exploitation is prevented after patching.
In contrast, security practice relies on a proof-of-concept exploit (PoC), that is an actual attack input or program that demonstrates end-to-end exploitability.
This gap motivates a benchmark that measures whether exploitation is truly blocked, rather than merely confirming that code or tests still execute.

To fill this gap,  we present \textbf{\bench}, a \emph{PoC-grounded} benchmark for real CVE repair. Rather than judging success with unit tests or a PoV, \bench{} aligns evaluation with security practice: a patch is successful only when the original PoC fails against the repaired program. We implement an end-to-end automated pipeline that generates candidate patches, executes the PoC, and performs a containerized two-environment differential check so that the only changing factor is the patch. \bench{} currently covers 23 real CVEs from 2017 to 2024 across nine common vulnerability categories, with task difficulty ranging from single-line edits to cross-file logical refactoring. The design favors rigor and reproducibility over scale, while remaining straightforward to extend.

Using \textbf{\bench}, we conduct a controlled study of 12 mainstream LLMs under a unified prompt and runtime.
Under the strict criterion that success requires the \emph{original PoC to fail} on the patched build, the strongest model repairs 5/23 (\(\approx\) 21.7\%), while several models repair at most one case; the average end-to-end success is about 7\%.
Most failures arise from incorrect localization (commonly 60–78\% of attempts), followed by unusable/incorrect diffs and logically incomplete fixes.
Some model families frequently abstain by declaring no vulnerability, reflecting a conservative bias that reduces false positives but misses real bugs.
We also evaluate two forms of assistance: (i) lightweight prompt hints (e.g., revealing the vulnerability type) and (ii) an \emph{agentic} workflow (multi-round analysis–patch–feedback).
Hints bring only marginal, inconsistent gains; the agent reduces abstentions and yields \emph{more} applicable patches, yet the PoC-validated end-to-end success remains low.
Our work has the following contributions:
\begin{itemize}
  \item \textbf{Benchmark.} \bench{} is a PoC-driven benchmark of real CVE repairs for Python, covering 23 CVEs across \textbf{9} categories. A repair counts as success only when the original PoC no longer works after patching.
  \item \textbf{Evaluation protocol.}  We design an end-to-end, reproducible pipeline (patch generation $\rightarrow$ PoC execution $\rightarrow$ dual-container differential validation) that enforces strict controls.
  \item \textbf{Systematic evaluation.} Under a unified prompt and runtime, we evaluate 12 LLMs and find consistently low end-to-end repair on real vulnerabilities, primarily due to mislocalization and unusable or incomplete patches. Prompt hints offer limited benefits; an agentic workflow improves intermediate signals but does not close the end-to-end gap.
  \item \textbf{Implications.} Aligning success with \emph{actual attack attempts} corrects the misconception that ``passing tests implies a fix'', and provides a rigorous foundation for future research on LLM-based vulnerability repair. We will release scripts and artifacts to reproduce our pipeline upon publication.
\end{itemize}

	\section{Related Work}
	LLM-based program repair has attracted increasing attention. We organize related work into five areas: general automatic program repair (APR) benchmarks, security-oriented repair benchmarks, PoC generation, vulnerability detection and reasoning, and general code generation benchmarks.

\noindent
\textbf{General APR Benchmarks.}
Defects4J~\cite{just2014defects4j} and BugsInPy~\cite{widyasari2020bugsinpy} focus on \emph{functional} bug repair for Java and Python, with developer-written unit tests as the oracle. SWE-Bench~\cite{jimenez2024swebench} raises the bar by validating patches under the full project CI test suite. While these benchmarks have advanced APR research, they do not target security vulnerabilities and therefore cannot assess whether LLMs can \emph{actually} fix exploitable flaws.

\noindent
\textbf{Security-Oriented Repair Benchmarks.}
To address this gap, researchers have built security-focused datasets, typically collecting real CVEs and their fixes. Large-scale corpora such as CVEfixes~\cite{bhandari2021cvefixes}, ReposVul~\cite{wang2024reposvul}, MegaVul~\cite{ni2024megavul}, and MoreFixes~\cite{akhoundali2024morefixes} provide cross-language pairs of vulnerabilities and fixes with metadata (e.g., CWE types), but most remain at the patch level and lack executable exploit validation.
Benchmarks closer to real exploitation include Vul4J~\cite{bui2022vul4j} and VUL4C~\cite{hu2025sok}, which incorporate PoV tests or exploit scripts to make evaluation more realistic. CVE-Bench~\cite{wang2025cvebench} further models black-box and white-box information layers for realistic bug reporting and integrates static analysis tools (e.g., Pylint, Bandit) to assess repair under varying information availability. Overall, the trend is moving from patch-level checks toward \emph{execution-driven} evaluation, yet systematic use of \emph{PoC failure as the sole success criterion} remains scarce, especially in the Python ecosystem.

\noindent
\textbf{PoC Generation.}
Effective repair depends on credible exploit validation. A complementary line of work aims to \emph{generate} PoCs automatically~\cite{simsek2025pocgen,long2023pocselfgen}. PoCGen~\cite{simsek2025pocgen}, for instance, combines LLMs with static/dynamic analysis to generate and validate PoCs in the npm ecosystem, outperforming template- or purely symbolic-execution-based approaches. Such methods are synergistic with PoC-driven repair benchmarks: reliable PoCs enable meaningful verification of whether a patch actually blocks exploitation.

\noindent
\textbf{Vulnerability Detection and Reasoning.}
Before repair, accurate detection and localization are crucial. SECVulEval~\cite{ahmed2025secvuleval} and recent analyses of LLM robustness~\cite{wang-etal-2023-recode, orvalho2025large} show that small changes in variable names, formatting, or library calls can derail model judgments, indicating insufficient robustness in security-critical settings. Reinforcement-learning-enhanced approaches such as SWE-RL~\cite{wei2025swerl} attempt to improve reasoning for software evolution tasks, but their impact on \emph{exploit-blocking} repairs remains to be fully assessed.

\noindent
\textbf{General Code Generation Benchmarks.}
HumanEval and MBPP are widely used to measure code generation. BigCodeBench~\cite{zhuo2024bigcodebench} emphasizes complex instructions and diverse library calls, pushing execution-oriented evaluation. However, these datasets lack explicit \emph{security semantics} and therefore cannot gauge whether generated code \emph{removes} exploitable behaviors. %

\begin{table*}[]
  \centering
  \caption{Representative benchmarks compared with \bench. ``PoV'' denotes proof-of-vulnerability (a minimal trigger), whereas \bench{} deems a fix successful only if the \emph{original PoC exploit} fails on the patched system.}
  \small
  \begin{tabular}{p{3.1cm}p{3.1cm}p{2.4cm}p{2.0cm}p{1cm}p{3.5cm}}
    \hline
    \textbf{Benchmark / Dataset} & \textbf{Objective} & \textbf{Validation} & \textbf{Language} & \textbf{Security?} & \textbf{Limitations} \\
    \hline
    HumanEval / MBPP & Code generation & Unit tests & Python & No & Synthetic tasks; not real vulnerabilities \\
    BugsInPy~\cite{widyasari2020bugsinpy} & Real bug repair & Test cases & Python & No & Functional bugs without security semantics \\
    Defects4J~\cite{just2014defects4j} & Java bug repair & Unit tests & Java & No & Software defects without security semantics \\
    SWE-Bench~\cite{jimenez2024swebench} & Software bug repair & Full-project CI test suite & Python & No & No security vulns; no PoC-based validation \\
    Vul4J~\cite{bui2022vul4j} & Java vulnerability repair & PoV tests & Java & Yes & PoV insufficient to prove attack blocking \\
    VUL4C~\cite{hu2025sok} & C/C++ vulnerability dataset & Exploit triggers / code analysis & C/C++ & Yes & Partial PoCs; not systematized for validation \\
    SECVulEval~\cite{ahmed2025secvuleval} & Vulnerability \emph{detection} benchmark & Function/statement labels & C/C++ & Yes & Focuses on detection; not repair evaluation \\
    CVE-Bench~\cite{wang2025cvebench} & CVE-centric repair evaluation & Test cases / fix commits & Multi-language & Yes & Limited validation; lacks PoC-driven criterion \\
    PoCGen~\cite{simsek2025pocgen} & Automatic PoC generation & --- & JavaScript & Yes & Targets PoC generation; not repair \\
    BigCodeBench~\cite{zhuo2024bigcodebench} & Code generation & Unit tests & Multi-language & No & Not security-focused; emphasizes code understanding \\
    \hline
    \textbf{\bench{} (ours)} & \textbf{Real vulnerability repair} & \textbf{PoC-driven validation} & \textbf{Python} & \textbf{Yes} & \textbf{Strict and reproducible} \\
    \hline
  \end{tabular}
  \label{tab:compare}
\end{table*}

The field has progressed from general bug repair toward security-oriented evaluation and has begun to incorporate exploit validation, layered information, and richer context. Nevertheless, three gaps remain: (i) Insufficient validation: success is often judged by compilation or unit tests rather than \emph{PoC/exploit}-level evidence; (ii) Limited realism: many inputs are function- or patch-level, with limited modeling of real reporting conditions (black-box vs.\ white-box) and deployment/dependency constraints; and (iii) Restricted coverage: many datasets are language- or vulnerability-specific, complicating systematic assessment across settings. 

\bench{} addresses these gaps with a PoC-driven success criterion, multi-level inputs (black-/white-box scenarios), and a containerized, automated environment for reproducible, end-to-end evaluation. Building on this trajectory, \bench{} offers a stricter and more practical benchmark for studying LLMs' capacity to repair security vulnerabilities.
	
	\section{Data Construction}
	\label{sec:data_construction}
	Our raw data is from the industrial partner that are strictly audited by the security experts. Since our objective is to construct a benchmark with Proof-of-Concept (PoC) rather than test cases, we focused on CVEs related to Python, as its environment setup requires comparatively less effort. 
Our initial dataset consists of 449 CVEs sourced from 192 open-source projects on GitHub. Since a single CVE can be addressed by multiple commits, particularly across different development branches, this corpus of vulnerabilities corresponds to a total of 862 security-fixing commits for our analysis. Git employs a prefix-matching mechanism that allows commits to be referenced by an abbreviated SHA-1 hash, so long as the prefix is unique within the repository. This functionality extends to platforms like GitHub, which can lead to the collection of variable-length commit hashes in practice. To ensure data consistency and establish a canonical representation for each vulnerability fix, we implemented a filtering protocol to retain only the longest available hash for each commit, thereby resolving any ambiguities from abbreviated forms. This refinement process resulted in a final dataset of 448 CVEs across 164 projects, corresponding to 684 unique commits.

To enrich our dataset, we sought to identify publicly available and executable PoC code for each CVE. Our primary source for this information was the reference links provided by the National Vulnerability Database (NVD) \cite{site:NVD}, which often led to advisories, analyses, or exploits. An initial automated crawl of these NVD references yielded 2,493 candidate URLs. Due to the infeasibility of manually verifying this large volume of potential PoCs, we developed and employed an automated verification pipeline based on the LLM.

Our process for analyzing URLs to identify executable PoC code is a multi-stage pipeline.
First, to capture content from dynamic web applications, we utilized an automated headless browser to visit each link. This step ensures that the final, client-side rendered HTML is fully loaded and available for analysis, rather than just the initial static source code.
Second, the fully rendered HTML is converted into Markdown format. This preprocessing step is critical for two reasons: 1) it substantially reduces the text volume, ensuring the input does not exceed the context token limit of the LLM, and 2) it distills the complex Document Object Model (DOM) hierarchy of HTML into a structured plain-text format. This transformation retains the essential layout and contextual relationships of the page's content, thereby enabling an effective structural analysis.
Finally, this processed Markdown text, along with a carefully designed prompt, is submitted to the LLM. The model is tasked with classifying the content of each page into one of the following three categories based on the nature of the PoC information provided:

\begin{enumerate}
    \item \textbf{Executable:} Contains complete, directly runnable code sufficient to reproduce the vulnerability.
    \item \textbf{Descriptive:} Lacks a complete script but provides a detailed natural language description of the exploit mechanism, often with partial code snippets, from which a full PoC could be constructed.
    \item \textbf{Brief:} Offers only a high-level summary of the vulnerability, without any actionable exploitation details or relevant code.
\end{enumerate}

\begin{figure}[htbp]
    \centering
    \includegraphics[width=0.8\linewidth]{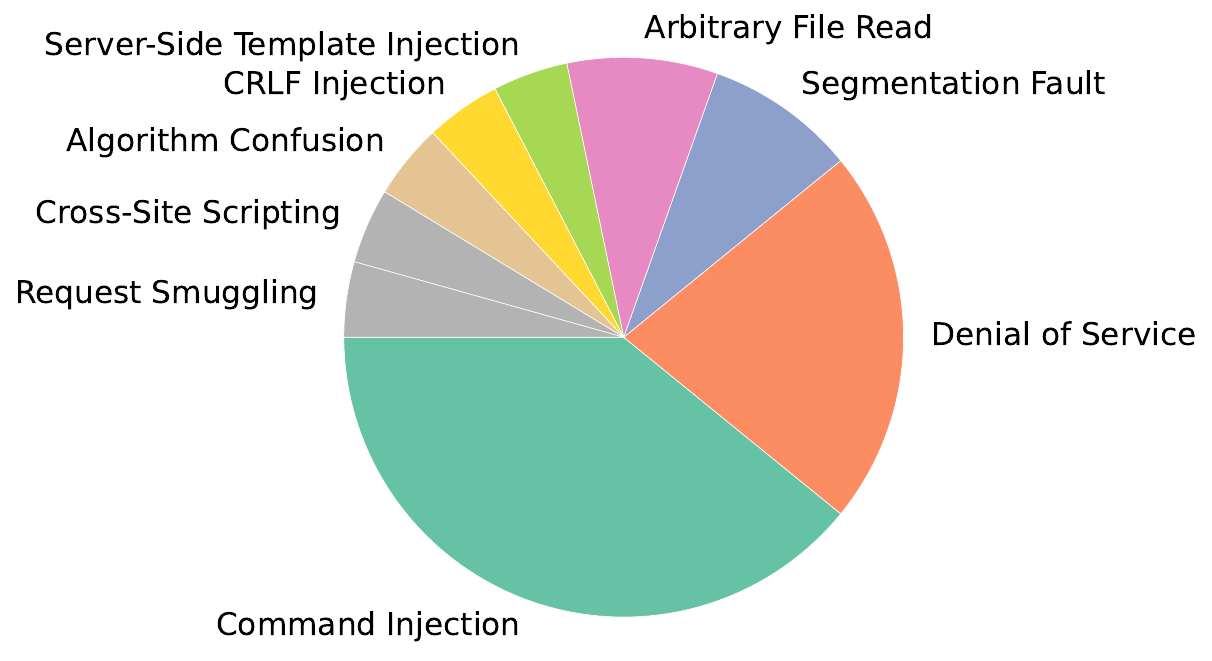}
    \caption{Distribution of vulnerability types of \bench.}
    \label{fig:vuln-types}
\end{figure}

We used the DeepSeek-V3 model \cite{liu2024deepseek} to analyze and classify the content of each reference link, and identified over 120 web links, corresponding to 109 unique CVEs, that were classified as containing executable PoCs. Subsequently, we conducted a verification of each of these LLM-flagged links to confirm the presence and functionality of the exploit code. 
Specifically, we constructed a PoC exploit based on the code provided in its corresponding reference materials for each CVE. We then executed each PoC against the vulnerable target project and validated its effectiveness by confirming that it produced the expected outcome.
This validation process affirmed the high precision of our LLM-based approach, revealing only a single false positive among all the candidates identified by the model.
Finally, the benchmark itself was curated to include 23 distinct CVEs from the years 2017 to 2024, representing nine different vulnerability types. The distribution of vulnerability types is shown in Figure \ref{fig:vuln-types}.

	\section{\bench}

\begin{figure*}[htbp]
    \centering
    \includegraphics[width=\linewidth]{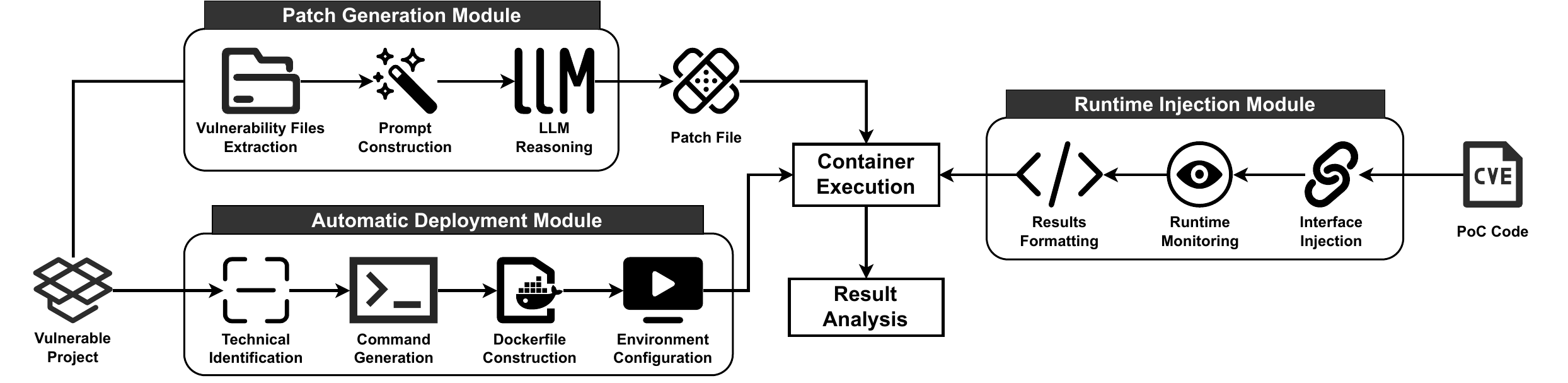}
    \caption{The overview of the \bench{} framework.}
    \label{fig:overview}
\end{figure*}

\noindent
\textbf{Overview.}
Section~\ref{sec:data_construction} shows the data construction process. This final benchmark comprises 23 CVEs representing nine distinct vulnerability classes, with repair complexities ranging from single-line fixes to complex algorithmic refactoring. To refine the task scope and mitigate the performance challenges of large context inputs, the methodology provides models only with the specific vulnerable files rather than entire project repositories. For evaluation, the framework employs a rigorous, automated pipeline that uses Docker to instantiate isolated environments for both the vulnerable and patched codebases, systematically executing the PoC in each to perform a differential analysis and automatically validate the patch's efficacy.
Our benchmark is architected around three core modules: patch generation, runtime injection, and automatic deployment. The overall architecture of this framework is shown in Figure \ref{fig:overview}.

\subsection{Patch Generation}

Analyzing entire software repositories for vulnerabilities is often infeasible for LLMs due to the immense scale of modern codebases, which typically exceeds the LLMs' operational context windows. To address this, our patch generation module employs a targeted approach, providing the model with only the specific source files containing a known vulnerability.
This methodology offers two primary advantages. First, it strategically refines the task from broad, unscoped vulnerability discovery to localized patch generation for a pre-identified flaw. Second, it not only mitigates the challenge of maintaining long-term memory across extensive inputs of LLMs but also avoids as much as possible their tendency to prioritize recently processed information, which can lead to over-focusing on a single code segment \cite{deng2024pentestgpt}.

To precisely identify the set of source files affected by a vulnerability, we perform a multi-stage analysis of the associated security patch commits. First, acknowledging that fixes can exist across multiple development branches, we constrain our analysis to the project's primary branch (e.g., main or master) to ensure a consistent and representative baseline. Then, we parse the file modifications within these commits to distinguish between substantive code changes and non-functional updates. Files related solely to metadata, such as dependency version bumps, documentation, or test case modifications, are filtered out. The remaining set of files is consequently identified as the ground truth for the location of the vulnerable code. 

To generate a patch, we construct the prompt as shown in Appendix~\ref{sec:prompt_appendix} for the LLMs containing the path and full source code of the file identified as vulnerable. The model is then tasked with a complex, two-part objective: first, it must autonomously localize the vulnerability within the provided code, and second, it must generate a syntactically correct patch in a standard diff format that remediates the flaw.
The complexity of this task is therefore twofold, simultaneously testing the model's capabilities in both deep semantic code understanding and precise, structured output generation. This dual requirement presents a considerable challenge for current models, moving far beyond simple code completion or modification.

\subsection{Runtime Injection}

To reliably verify vulnerabilities, effective PoC code is essential. However, the diversity of vulnerability types and exploitation methods leads to significant inconsistencies across PoCs, making standardized, automated execution difficult. To normalize this process, our framework employs a runtime injection module that acts as a standardized wrapper to create a unified execution and analysis interface. This module is composed of three core components: \textit{Interface Injection}, \textit{Runtime Monitoring}, and \textit{Result Formatting}. This modular approach allows our system to manage and evaluate diverse PoCs in a consistent and repeatable manner.

\textbf{Interface Injection:} To ensure reliable execution and consistent state monitoring, our module injects a unified control interface into each PoC through a controlled runtime environment. First, the PoC is launched as a child process using Python's subprocess module, which allows our framework to capture and redirect its standard input, output, and error streams. Furthermore, before execution, our interface completes the necessary environment setup by automatically installing all of the PoC's required runtime dependencies. This approach of I/O redirection and dependency provisioning guarantees that the PoC can operate as intended and successfully trigger the target vulnerability for analysis.

\textbf{Runtime Monitoring:} During the execution of each PoC, our framework systematically monitors and records key runtime information, including complete input and output streams,  as well as critical runtime metrics such as its total execution time, any generated error messages, and detailed crash reports or non-zero exit codes. The collected data is then analyzed and validated to determine whether it matches the result expected to provide a conclusive determination of whether the PoC successfully exploits the target vulnerability.

\textbf{Result Formatting:} For subsequent processing and systematic analysis, our module formats and exports all captured PoC runtime data. To ensure no details are omitted during this process, all input and output data are uniformly encoded, preserving a full and lossless record of the PoC’s execution for future analysis.

\subsection{Automatic Deployment}

To establish the precise, vulnerable state for each project sourced from GitHub, we leverage its version control history. For each CVE, we first identify the official security-fixing commit. The direct parent of this patch commit logically represents the final state of the codebase containing the vulnerability before remediation. We then check out this specific parent commit hash. This procedure ensures that the environment under analysis is configured with the exact vulnerable code, thereby creating a verifiable and accurate baseline against which patch efficacy is measured.

The projects targeted in our study are predominantly developed in Python, a language known for its extensive and varied ecosystem. This diversity, however, introduces considerable challenges in establishing reproducible runtime environments. 
Projects frequently rely on disparate dependency management systems, such as pip with \texttt{requirements.txt} files, Poetry, or Conda environments, and often require bespoke installation scripts.
To address this challenge and streamline the setup process, we have introduced an automated deployment module. This component is designed to identify the specific technical stack and configuration requirements of a given project, thereby ensuring a consistent, reliable, and fully automated deployment for vulnerability analysis.

The initial phase of automated deployment involves a static analysis of the project's source code to identify its dependencies. %
This process must account for the evolution of Python packaging standards, from legacy \texttt{setup.py} scripts that rely on distutils and setuptools to the modern \texttt{pyproject.toml} file~\cite{site:pep518}, which has become the standard for contemporary projects.
To accurately resolve the project's complete dependency graph, our deployment module employs a prioritized parsing strategy. It preferentially analyzes \texttt{pyproject.toml } due to its standardized and comprehensive nature. If this file is not present, the module falls back to parsing \texttt{setup.py}, and subsequently, \texttt{requirements.txt}, to ensure all necessary dependencies are captured. Upon successful resolution, the module automatically generates the requisite shell commands to invoke the appropriate package manager, thereby systematically reconstructing the project's runtime environment.

To guarantee a rigorous and isolated testing environment, our framework leverages Docker to containerize the deployment and verification process. For each vulnerable project, a custom Dockerfile is generated. This file codifies the previously resolved dependencies and system requirements into a reproducible build specification, serving as a blueprint for the runtime environment.
From this specification, the framework provisions two parallel container instances for each run: a vulnerable instance executing the original, unmodified code, and a patched instance executing the version with the generated patch applied. This architecture is fundamental to performing a differential analysis, wherein the same PoC exploit is executed under identical conditions across both containers. The strict isolation afforded by containerization ensures that the sole variable between tests is the presence of the patch, thereby preventing any cross-contamination of state and validating that the outcome is a direct consequence of the repair attempt. Considering the ethical reasons, we do not make PoCs publicly available, but with limited research access.

	\section{Evaluation}

\subsection{Evaluation Setting}

To demonstrate the benchmark's effectiveness and establish initial baselines, we conducted a large-scale comparative study of 12 leading LLMs. 
This evaluation covered both proprietary and open-source models. On the proprietary side, we included Gemini 2.5 Pro \cite{comanici2025gemini}, Gemini 2.5 Flash, and Gemini 2.0 Flash, as well as GPT-4o \cite{hurst2024gpt}, GPT-o4 mini \cite{site:o4-mini}, and GPT-3.5 Turbo \cite{site:gpt-3.5}. The open-source set comprised DeepSeek V3 671B \cite{liu2024deepseek}, DeepSeek R1 671B \cite{guo2025deepseek}, and four Qwen-3 models \cite{yang2025qwen3}—the 8B and 235B variants, each available in both base and Thinking versions. We also  include SWE-Agent~\cite{yang2024sweagent}.
Each model was tasked with patching the same set of vulnerabilities under uniform conditions to ensure a controlled and fair comparison. 
The experimental code is released publicly as a robust instrument to drive further advancements and facilitate standardized evaluation of vulnerability fixing \cite{site:vulbench}.

\subsection{Evaluation Metrics}
\label{sec:evalm}
To comprehensively evaluate the vulnerability repair capabilities of LLMs, we designed a hierarchical evaluation metric. This metric is built upon the cornerstone of a primary success criterion: (1) whether the patch can successfully neutralize a real-world exploit and (2) is supplemented by a series of diagnostic metrics intended to provide a deep analysis of each stage, from vulnerability localization to patch generation.

\noindent
\textbf{Primary metric: PoC-validated repair success (\(P_{\text{succ}}\)).}
Our main metric is the end-to-end ability of a patch to neutralize the exploit. For each CVE, we execute the PoC on the unpatched baseline (must succeed) and on the patched build (must fail). A repair is counted as successful only if:
(i) the model emits a patch that applies to the target project (clean or fuzzy apply; see below),
and (ii) the PoC that succeeds on the baseline \emph{fails} after patching.
We report \(P_{\text{succ}}\) as the proportion of CVEs with PoC-validated fixes.

\noindent
\textbf{Secondary diagnostics (for analysis, not primary).}
Format correctness (\(P_{\text{corr}}\)) is to reflect whether a generated unified diff can be applied to the codebase cleanly without any offset or fuzzy matching. 
Our patch application process employs a two-stage strategy. The primary method utilizes \texttt{git apply} \cite{gitapplydocs} for precise patch integration. If this strict application fails, the system will attempt to use the \texttt{patch} \cite{ibmpatchdocs} command to reapply the patch. The patch command is employed as a secondary step due to its support for context-based fuzzy matching, which can successfully apply patches by tolerating minor line number offsets in the target file.
A clean apply is marked \ding{51} for one patch. To tolerate minor LLM offsets, patches that \emph{modify the intended file(s)} only via fuzzy apply are denoted \ding{55}.  
We evaluate each generated patch against two pri-
mary criteria: 
[F]ormat Correctness, which assesses whether the patch can be applied to the codebase successfully, and [R]epair Efficacy, which determines if the patch successfully remediates the vulnerability.

\noindent
\textbf{Detection abstention (\(V_{\text{dnf}}\)).} The fraction of CVEs where the model claims the code is safe or emits no patch (\emph{Did Not Find}). This captures conservative bias / recall loss at the localization stage. These diagnostics help attribute failures to \emph{localization} vs.\ \emph{generation/applicability}.

\noindent
\textbf{Reporting convention.}
All results are reported first and foremost with \(P_{\text{succ}}\) (per-model success rate and per-CVE outcomes). Then we provide \(P_{\text{corr}}\) and \(V_{\text{dnf}}\) to analyze failure modes. 

\noindent
\textbf{Composite ranking score $S_p$ (for leaderboard compactness).}
Our primary evaluation metric is \(P_{\text{succ}}\) (PoC-validated repair). However, in a security setting with a modest number of CVEs and generally low success rates, different models often share similar or identical \(P_{\text{succ}}\). This creates two practical issues for reporting: (i) ties and near-ties that obscure relative ordering, and (ii) lack of visibility into \emph{how} a model fails (abstention vs.\ format). To provide a \emph{compact, tie-breaking} summary while keeping \(P_{\text{succ}}\) as the scientific endpoint, we introduce a lightweight composite score that (a) \emph{prioritizes} end-to-end repair, (b) \emph{nudges} models with better patch applicability, and (c) \emph{penalizes} excessive abstention (\(V_{\text{dnf}}\)). This score (denoted as $S_p$) is used only for leaderboard ordering and not for hypothesis testing.

\noindent
\textbf{Definition of $S_p$:}
\begin{equation}
\label{eq:score_metric}
S_{p} \;=\;
\frac{(1+\beta^2)\,\big(P_{\text{amend}} \cdot P_{\text{succ}}\big)}
{\beta^2 \cdot P_{\text{amend}} + P_{\text{succ}}}
\;\cdot\; \left(1 - 0.5 \cdot V_{\text{dnf}}\right)
\end{equation}

where $P_{\text{amend}} \;=\; \ln\!\left(P_{\text{corr}} + 1\right)$,
\(P_{\text{corr}}\) is the proportion of format-correct patches (clean apply), \(P_{\text{succ}}\) is the PoC-validated repair rate, and \(V_{\text{dnf}}\) is the abstention rate.
We adopt an F\(_\beta\)-style core because it is a principled way to combine two dimensions with unequal importance. We use the common setting of $F_{beta}$ with \(\beta{=}2\) so that increases in \(P_{\text{succ}}\) dominate the score relative to \(P_{\text{corr}}\) (successful defense is the end goal). The logarithmic transform \(P_{\text{amend}}\) encodes diminishing returns: once patches are broadly applicable, further formatting gains should not outweigh end-to-end security impact. Finally, the multiplicative penalty \(\left(1 - 0.5 \cdot V_{\text{dnf}}\right)\) scales the score by localization recall without erasing hard-won successes (even \(V_{\text{dnf}}{=}1\) halves but does not zero-out the score).

\textbf{Intended use and properties.}
The composite is \emph{monotone} in \(P_{\text{succ}}\) (higher repair \(\Rightarrow\) higher score), gives smaller, diminishing credit to \(P_{\text{corr}}\) when \(P_{\text{succ}}\) is fixed, and decreases with \(V_{\text{dnf}}\). By construction it remains in \([0,1]\). Importantly, this composite does \textbf{not} replace \(P_{\text{succ}}\); all scientific conclusions are drawn from \(P_{\text{succ}}\) (primary), with \(P_{\text{corr}}\) and \(V_{\text{dnf}}\) serving as diagnostics. The composite only resolves leaderboard ties and summarizes secondary behaviors.

\noindent
\textbf{Practical notes.}
A patch with correct format counted as cleanly applied (\ding{51}) contributes to \(P_{\text{corr}}\), and it contributes to \(P_{\text{succ}}\) only when the PoC fails post-patch. If no patch is emitted, the trial contributes to \(V_{\text{dnf}}\). This separation ensures that the \emph{primary endpoint} remains PoC-validated efficacy, while secondary diagnostics support root-cause analysis.

\subsection{Research Questions (RQs)}
Based on our PoC validated setup on \emph{\bench}, we formulate three questions that structure the study across baseline capability, failure diagnosis, and mitigation strategies.

\noindent
\textbf{[RQ1]} \textbf{PoC Validated Repair Capability.} \emph{How capable are LLMs at repairing real world vulnerabilities end to end under PoC validation?}
  We quantify end to end repair using \(P_{\text{succ}}\) (the PoC fails after patching) as the primary outcome, and use \(P_{\text{corr}}\) (the patch applies) and \(V_{\text{dnf}}\) (did not find, i.e., detection failure) as secondary diagnostics to dissect the pipeline.

\noindent
\textbf{[RQ2]} \textbf{Failure Analysis in LLM Based Patching.} \emph{Where does the LLM driven pipeline break, and what are the dominant root causes?}
  We construct a failure taxonomy over three stages: (i) detection and localization (\(V_{\text{dnf}}\)), (ii) patch synthesis and applicability (\(P_{\text{corr}}\)), and (iii) post patch efficacy measured by the PoC outcome, and we relate failure modes to vulnerability difficulty.

\noindent
\textbf{[RQ3]} \textbf{Efficacy of Prompting and Agentic Augmentation.} \emph{Do advanced input and interaction strategies overcome the limitations observed with simpler single pass repairs?}
  We evaluate two families of interventions, \emph{Static Prompt Enhancement} (adding targeted context cues and format scaffolds) and \emph{Dynamic Agentic Interaction} (multi step planning with tool use and iterative refinement), and we measure their impact on \(P_{\text{succ}}\) and on the failure mix identified in \textbf{RQ2}.

\noindent\textit{Rationale and flow.}
\textbf{RQ1} establishes a PoC grounded baseline across models under a unified runtime and prompt.
\textbf{RQ2} explains low end to end rates by attributing errors to specific pipeline stages using \(P_{\text{corr}}\) and \(V_{\text{dnf}}\).
\textbf{RQ3} then tests whether stronger inputs or interaction can lift \(P_{\text{succ}}\) or shift failure mass away from detection and synthesis bottlenecks.

\subsection{Implementation}
Our benchmark is developed in Python, initiates its process by reconstructing the vulnerable state of a target project. This is accomplished by reverting the project's repository to the parent commit of the official repair patch. Subsequently, the vulnerable source files are utilized to construct prompts for the LLMs, which are queried via an API to identify potential vulnerabilities and generate a corresponding patch. For evaluation, \bench{} automates the deployment of the vulnerable project within Docker containers. 
The efficacy of the generated patch is then assessed by executing a PoC against two container instances: one with the patch applied and one without. Through a comparative analysis of their runtime status and output, the system determines the patch's effectiveness. Additionally, the framework verifies the successful application of the patch to ensure the integrity of its format and content.
To ensure a fair and reproducible evaluation of the LLMs' patch generation performance, all experiments were conducted in a consistent environment. The experiments were performed on a server equipped with two Intel(R) Xeon(R) E5-2630v4 CPUs and 128GB of RAM running Ubuntu 20.04.

	\section{Results and Analysis}
	\subsection{RQ1: PoC-Validated Repair Capability}

\paragraph{Evaluation Setup.}
To ensure clarity and reproducibility in answering RQ1, a rigorous experimental protocol was established.
\begin{enumerate}
    \item \textbf{Models under test.} We evaluate 12 leading LLMs. This diversity enables thorough comparison across architectural philosophies, thinking and model scales.
    \item \textbf{Benchmark and task.} For a curated set of real-world \bench{} CVEs, each model must produce a fixing patch by (i) localizing the vulnerability in the provided source files and (ii) generating a syntactically correct patch in the standard diff format.
\end{enumerate}

\noindent
\textbf{End-to-end repair (\(P_{\text{succ}}\) \& $S_p$)}
An analysis of the patch efficacy results in Table~\ref{table:overall_result}. \(P_{\text{succ}}\) is our main evaluation metric and $S_p$ is the complementary metrics as illustrated at \textit{Primary metric: PoC-validated repair success} and \textit{Composite ranking score $S_p$} in Section~\ref{sec:evalm}. Table~\ref{table:overall_result} reveals that the overall repair success rates are not so good, with the top-performing model, Gemini 2.5 Pro, achieving a repair efficacy of 0.217, which translates to a final score of 0.226. The vast majority of models scored below 0.20, with several scoring zero. Moreover, we conducted repeated experiments for verification, and the results showed pretty small variation. This data clearly indicates a significant ``capability gap'' when LLMs are tasked with security-critical repair tasks.
Specifically, Gemini 2.5 Pro stands alone at the top, successfully fixing 5 distinct CVEs. In stark contrast, multiple models, including GPT 4o, Qwen3 8B, and Qwen3 235B, failed to produce a single successful fix, resulting in scores of zero. The models at the top of the leaderboard (e.g., Gemini 2.5 Pro, DeepSeek R1 671B) are recognized state-of-the-art models, while smaller or older-generation models cluster at the bottom. It can be found that fixing vulnerabilities is not a general skill of LLMs but a frontier capability, which in turn validates the high complexity of the task and its reliance on cutting-edge model architectures.

\noindent
\textbf{Patch synthesis (\(P_{\text{corr}}\)).}
The \bench{} framework was utilized to systematically analyze and score the efficacy of each submitted patch. The aggregate results are summarized in Table~\ref{table:all}. We use the evaluation metrics as shown in \textit{Secondary diagnostics} of Section~\ref{sec:evalm}.
A check mark (\ding{51}) in the [R]epair column indicates a successful repair, verified by the failure of the PoC exploit against the patched code. For format correctness, a check mark (\ding{51}) in the [F]ormat column signifies a patch that applies cleanly. However, to accommodate minor LLM-generated inaccuracies like incorrect line numbers, our framework supports context-based fuzzy matching with a certain degree of offset. A patch that is successfully applied but relies on this fuzzy matching is denoted by an error mark (\ding{55}), indicating it modified the project despite minor formatting offsets.
Conditional on attempting a fix, many failures arise from \emph{diff conformance} rather than security logic. In Table~\ref{table:all}, [F] \ding{51} denotes clean apply, while \ding{55} denotes fuzzy-apply via context-offset matching. When formatting defects (e.g., missing line numbers/context or malformed headers) prevent application, the attempt fails regardless of localization quality—low \(P_{\text{corr}}\) is a practical bottleneck, and GPT-family models often produce semantically relevant but syntactically broken patches that even fuzzy matching cannot apply.

\noindent
\textbf{Detection (\(V_{\text{dnf}}\)).}
A striking finding in Table~\ref{table:overall_result} is the exceptionally high rate of ``No Vulnerabilities Found''. DeepSeek V3 671B exhibited this behavior in 87.0\% of cases, while the GPT series ranged between 43.5\% and 56.5\%.
This phenomenon is not a simple failure to generate; it is a failure to detect and a refusal to act. This ``conservative bias'' indicates that when faced with uncertainty, these models are optimized to avoid false positives (incorrectly flagging safe code) at the cost of producing a high number of false negatives (missing real vulnerabilities). This is a deeply perilous trait as it creates a false sense of security. This behavior is likely a result of model training objectives or alignment processes that excessively penalize the generation of ``wrong'' or ``harmful'' code, leading models to learn that the safest strategy is to claim ignorance or declare the input safe.

\begin{table*}[!ht]
    \centering
    \caption{Performance of LLMs on \bench{} (T means thinking model).}
    \begin{adjustbox}{max width=\textwidth, max height=\textheight, keepaspectratio}
    \begin{tabular}{c|cc|cc|cc|cc|cc|cc|cc|cc|cc|cc|cc|cc}
    \Xhline{1.5pt}
        \multirow{2}{*}{\textbf{CVE ID}} 
        & \multicolumn{2}{c|}{\textbf{\shortstack{Gemini \\ 2.5 Pro}}} 
        & \multicolumn{2}{c|}{\textbf{\shortstack{Gemini \\ 2.5 Flash}}} 
        & \multicolumn{2}{c|}{\textbf{\shortstack{Gemini \\ 2.0 Flash}}} 
        & \multicolumn{2}{c|}{\textbf{\shortstack{GPT \\ 4o}}} 
        & \multicolumn{2}{c|}{\textbf{\shortstack{GPT \\ o4 mini}}} 
        & \multicolumn{2}{c|}{\textbf{\shortstack{GPT \\ 3.5 Turbo}}} 
        & \multicolumn{2}{c|}{\textbf{\shortstack{DeepSeek \\ V3 671B}}} 
        & \multicolumn{2}{c|}{\textbf{\shortstack{DeepSeek \\ R1 671B}}} 
        & \multicolumn{2}{c|}{\textbf{\shortstack{Qwen3 \\ 8B}}} 
        & \multicolumn{2}{c|}{\textbf{\shortstack{Qwen3 \\ 235B}}} 
        & \multicolumn{2}{c|}{\textbf{\shortstack{Qwen3\\ 8B T}}} 
        & \multicolumn{2}{c}{\textbf{\shortstack{Qwen3\\ 235B T}}} \\ 
        \cline{2-25}
        & F & R & F & R & F & R & F & R & F & R & F & R & F & R & F & R & F & R & F & R & F & R & F & R \\ \hline
        2017-16615  & \ding{51} & \ding{51} & ~ &   & \ding{51} &   & \ding{55} &   & ~ &   & ~ &   & \ding{51} & \ding{51} & \ding{55} & \ding{51} & \ding{55} & ~ &   & ~ & \ding{55} & ~ & \ding{51} & \ding{51} \\ 
        2017-16618  & \ding{51} & \ding{51} & \ding{51} & \ding{51} & \ding{55} &   & \ding{55} &   & ~ &   & ~ &   & \ding{55} & \ding{51} & ~ &   &   & ~ &   & ~ & \ding{55} & ~ & ~ &   \\ 
        2017-2809  & \ding{51} & \ding{51} & ~ &   & \ding{51} & \ding{51} & \ding{55} &   & ~ &   & \ding{55} & \ding{51} & \ding{55} & \ding{51} & \ding{51} & \ding{51} & \ding{51} & ~ &   & ~ &   & ~ & \ding{51} & \ding{51} \\ 
        2018-15560  & ~ &   & ~ &   & ~ &   & ~ &   & ~ &   & ~ &   & ~ &   & \ding{55} & \ding{51} &   & ~ &   & ~ & \ding{55} & ~ & \ding{51} & ~ \\ 
        2019-10800  & \ding{55} & \ding{51} & \ding{55} & \ding{51} & \ding{51} &   & \ding{55} &   & \ding{55} & \ding{51} & ~ &   & ~ &   & ~ &   &   & ~ & \ding{55} & ~ & \ding{55} & ~ & ~ &   \\ 
        2020-10109  & \ding{51} &   & \ding{51} &   & \ding{51} & \ding{51} & ~ &   & ~ &   & ~ &   & ~ &   & ~ &   & \ding{55} & ~ & \ding{55} & ~ &   & ~ & \ding{55} &   \\ 
        2021-23980  & ~ &   & ~ &   & ~ &   & ~ &   & ~ &   & ~ &   & ~ &   & ~ &   & \ding{55} & ~ &   & ~ &   & ~ & \ding{55} &   \\ 
        2021-29063  & \ding{51} &   & \ding{51} &   & \ding{55} &   & ~ &   & ~ &   & ~ &   & ~ &   & ~ &   &   & ~ &   & ~ &   & ~ & ~ &   \\ 
        2021-32837  & \ding{55} &   & ~ &   & \ding{51} &   & ~ &   & ~ &   & ~ &   & ~ &   & \ding{55} &   & \ding{55} & ~ & \ding{51} & ~ & \ding{55} & ~ & ~ &   \\ 
        2022-21797  & \ding{51} & \ding{51} & ~ &   & \ding{55} &   & ~ &   & ~ &   & ~ &   & ~ &   & \ding{55} &   & \ding{55} & ~ &   & ~ &   & ~ & \ding{55} &   \\ 
        2022-24065  & ~ &   & \ding{51} & ~ & \ding{51} &   & ~ &   & ~ &   & \ding{55} &   & ~ &   & \ding{51} & \ding{51} &   & ~ & \ding{55} & ~ & \ding{55} & ~ & \ding{55} & ~ \\ 
        2022-24439  & \ding{55} &   & \ding{55} &   & ~ &   & ~ &   & ~ &   & ~ &   & ~ &   & ~ &   & \ding{55} & ~ &   & ~ & \ding{55} & ~ & ~ &   \\ 
        2022-29217  & ~ &   & \ding{55} &   & \ding{55} &   & ~ &   & ~ &   & ~ &   & ~ &   & ~ &   & \ding{55} & ~ &   & ~ & \ding{55} & ~ & ~ &   \\ 
        2023-2356  & ~ &   & ~ &   & \ding{51} &   & ~ &   & ~ &   & ~ &   & ~ &   & ~ &   & \ding{55} & ~ & \ding{55} & ~ &   & ~ & ~ &   \\ 
        2023-26145  & ~ &   & \ding{55} &   & \ding{51} &   & ~ &   & \ding{51} &   & \ding{55} &   & ~ &   & \ding{51} &   &   & ~ & \ding{55} & ~ & \ding{55} & ~ & ~ &   \\ 
        2023-32309  & ~ &   & ~ &   & ~ &   & ~ &   & ~ &   & \ding{55} &   & ~ &   & ~ &   &   & ~ &   & ~ &   & ~ & ~ &   \\ 
        2023-49081  & ~ &   & \ding{51} &   & \ding{51} &   & ~ &   & ~ &   & ~ &   & ~ &   & ~ &   & \ding{55} & ~ &   & ~ &   & ~ & ~ &   \\ 
        2023-49083  & \ding{55} &   & ~ &   & \ding{51} &   & ~ &   & ~ &   & ~ &   & ~ &   & ~ &   & \ding{55} & ~ &   & ~ &   & ~ & ~ &   \\ 
        2024-21503  & ~ &   & ~ &   & \ding{51} &   & ~ &   & ~ &   & \ding{55} &   & ~ &   & ~ &   & \ding{55} & ~ &   & ~ &   & ~ & ~ &   \\ 
        2024-26151  & \ding{51} &   & ~ &   & \ding{51} &   & ~ &   & ~ &   & ~ &   & ~ &   & ~ &   & \ding{55} & ~ & \ding{51} & ~ & \ding{55} & ~ & ~ &   \\ 
        2024-28102  & \ding{55} &   & ~ &   & \ding{51} &   & ~ &   & ~ &   & \ding{55} &   & ~ &   & \ding{51} &   &   & ~ & \ding{51} & ~ & \ding{55} & ~ & ~ &   \\ 
        2024-3571  & ~ &   & ~ &   & ~ &   & ~ &   & ~ &   & ~ &   & ~ &   & ~ &   & \ding{55} & ~ &   & ~ &   & ~ & ~ &   \\ 
        2024-4340  & \ding{55} &   & ~ &   & \ding{55} &   & ~ &   & ~ &   & ~ &   & ~ &   & ~ &   & \ding{55} & ~ &   & ~ &   & ~ & ~ &   \\ \Xhline{1.5pt}
    \end{tabular}
    \end{adjustbox}
    \label{table:all}
\end{table*}

\newcolumntype{M}[1]{>{\centering\arraybackslash}m{#1}}

\begin{table}[!ht]
    \centering
    \caption{The overall assessment result of LLMs' capabilities in automated vulnerability repair. T means thinking model and Non-T means non-thinking model.}
    \begin{tabular}{c M{1.8cm} cccc}
    \hline
        \textbf{Type} & \textbf{Model} & \textbf{$V_{dnf}$} & \textbf{$P_{corr}$} & \textbf{$P_{succ}$} & \textbf{$S_p$} \\ \hline
        \multirow{6}{*}{\shortstack{T}} 
        & Gemini 2.5 Pro      & 0.000 & 0.304 & 0.217 & \cellcolor{red!46}0.226 \\ 
        & Gemini 2.5 Flash    & 0.174 & 0.217 & 0.087 & \cellcolor{red!18}0.089 \\ 
        & GPT o4 mini     & 0.565 & 0.043 & 0.043 & \cellcolor{red!6}0.031 \\ 
        & DeepSeek R1 671B    & 0.217 & 0.174 & 0.174 & \cellcolor{red!30}0.152 \\ 
        & Qwen3 8B T          & 0.000 & 0.000 & 0.000 & \cellcolor{red!0}0.000 \\ 
        & Qwen3 235B T        & 0.130 & 0.130 & 0.087 & \cellcolor{red!18}0.086 \\ \hline
        \multirow{6}{*}{\shortstack{Non-T}} 
        & Gemini 2.0 Flash    & 0.000 & 0.565 & 0.087 & \cellcolor{red!20}0.104 \\ 
        & GPT 4o          & 0.478 & 0.000 & 0.000 & \cellcolor{red!0}0.000 \\ 
        & GPT 3.5 Turbo   & 0.435 & 0.000 & 0.043 & \cellcolor{red!0}0.000 \\ 
        & DeepSeek V3 671B    & 0.870 & 0.043 & 0.130 & \cellcolor{red!10}0.052 \\ 
        & Qwen3 8B            & 0.000 & 0.043 & 0.000 & \cellcolor{red!0}0.000 \\ 
        & Qwen3 235B          & 0.391 & 0.130 & 0.000 & \cellcolor{red!0}0.000 \\ \hline
    \end{tabular}
    \label{table:overall_result}
\end{table}

\noindent
\textbf{Factors: scale and explicit reasoning}
A direct comparison between ``Thinking'' and ``Non-Thinking'' variants of the same base models reveals significant performance differences. The ``Thinking'' version of Qwen3 235B achieved a repair efficacy $P_{succ}$ of 0.087, whereas its ``Non-Thinking'' counterpart scored 0. Similarly, the reasoning-focused DeepSeek R1 671B (score $S_p$ 0.152) significantly outperformed DeepSeek V3 671B (score 0.052). This phenomenon provides strong evidence that successful vulnerability repair is not a simple, single-step generation task but a complex problem that benefits from an explicit, structured reasoning process. ``Thinking'' models, by externalizing their chain of thought, are better able to decompose the problem and address each step, thereby increasing their probability of success. In contrast, ``Non-Thinking'' models appear to attempt an intuitive leap from problem to solution in a single forward pass, a strategy that is ill-suited to the complexity of vulnerability repair.

\paragraph{Answer to RQ1:}At present, LLMs generally perform poorly in generating patches for vulnerability repair; however, there is a clear trend that larger models with stronger reasoning capabilities tend to achieve better results.

\subsection{RQ2: Failure Analysis}

To investigate the root causes behind the low success rates of LLMs in automated vulnerability repair, based on the analysis of RQ1, we guess that a model performance is tied to the difficulty of the vulnerability itself. To validate this, we conducted a multi-level analysis and revealed the core challenges that current LLMs face in security repair tasks.

\noindent
\textbf{Vulnerability Complexity.}
To establish a reliable difficulty baseline, we first engaged expert vulnerability security analysts to manually classify each vulnerability in the dataset into three tiers based on the technical complexity required for discovery and exploitation: \texttt{Easy}, \texttt{Medium}, and \texttt{Hard}. 

Specifically,  \texttt{Easy} vulnerabilities are typically localized to a single high-risk function and can be remediated with minimal contextual analysis.  \texttt{Medium} vulnerabilities necessitate a broader understanding of control and data flow, as their resolution often requires localized code refactoring or the implementation of new logic. Finally,  \texttt{Hard} vulnerabilities are typically algorithmic or design-level flaws, such as cryptographic bypasses or subtle errors in protocol logic. Fixing these issues requires a deep comprehension of the application's architecture and may involve large-scale refactoring across multiple components.

We then conducted a targeted evaluation of six LLMs noted for their advanced reasoning capabilities, assessing their repair performance against these stratified difficulty levels. The detailed results are presented in Table \ref{table:vulnerability_levels}.

\begin{table}[!ht]
    \centering
    \caption{Number of vulnerabilities fixed by LLMs for different difficulty levels (T means thinking model).}
    \begin{tabular}{ccccc}
    \Xhline{1.5pt}
        ~ & \textbf{Easy} & \textbf{Medium} & \textbf{Hard} & \textbf{Total} \\ \hline
        Gemini 2.5 Pro & 3 & 2 & 0 & 5 \\ \
        Gemini 2.5 Flash & 1 & 1 & 0 & 2 \\ 
        GPT o4 mini & 0 & 1 & 0 & 1 \\ 
        DeepSeek R1 671B & 2 & 1 & 1 & 4 \\ 
        Qwen3 8B T & 0 & 0 & 0 & 0 \\ 
        Qwen3 235B T & 2 & 0 & 0 & 2 \\ \Xhline{1.5pt}
    \end{tabular}
    \label{table:vulnerability_levels}
\end{table}

The data indicates that LLMs are most effective when addressing vulnerabilities classified as \emph{Easy}. However, as the task difficulty increases, their performance declines rapidly. For instance, the top-performing model, Gemini 2.5 Pro, successfully repaired 3 \emph{Easy} and 2 \emph{Medium} vulnerabilities but failed on all \emph{Hard}. Similarly, most models had a zero success rate in the \emph{Hard} category. DeepSeek R1 671B was the sole model to fix a \emph{Hard} vulnerability, but this single success only serves to highlight the general rule: highly complex, real-world vulnerabilities remain beyond the reliable repair capabilities of LLMs.

\begin{table*}[]
    \centering
    \caption{An analysis of success and failure in LLMs vulnerability repair.}
    \begin{tabular}{ccccccc}
    \Xhline{1.5pt}
        ~ & \shortstack{Gemini 2.5\\Pro} & \shortstack{Gemini 2.5\\Flash} & \shortstack{GPT\\o4 mini}  & \shortstack{DeepSeek R1\\671B} & \shortstack{Qwen3\\8B Thinking} & \shortstack{Qwen3\\235B Thinking} \\ \hline
        Repair Success ($P_{succ}$)  & 0.217 & 0.087 & 0.043 & 0.174 & 0.000 & 0.087 \\ 
        Miss Vulnerable Code & 0.609 & 0.696 & 0.739 & 0.609 & 0.782 & 0.565 \\ 
        Patch Error & 0.174 & 0.217 & 0.217 & 0.217 & 0.217 & 0.348 \\ \Xhline{1.5pt}
    \end{tabular}
    \label{table:repair_analysis}
\end{table*}

\noindent
\textbf{Reasons for Failure.}
To further deconstruct the reasons for LLMs' failure to repair vulnerability,  we categorize unsuccessful repair attempts into two primary modes: \emph{Localization Failure}, where the model fails to identify the vulnerability in the code, and \emph{Generation Failure}, where the model attempts a fix but fails to produce a correct and effective patch. 
We systematically analyze the outputs and reasoning of all models, categorizing three outcomes: Repair Success $P_{succ}$, Miss Vulnerable Code, and Patch Error, as shown in Table \ref{table:repair_analysis}. It can be seen that the predominant failure is the inability of LLMs to localize the vulnerability, which accounts for the majority of unsuccessful repairs. Patch generation errors constitute a secondary but significant cause, representing approximately 20\% of all attempts.

\noindent
\textbf{Case evidence: applicability \& mitigation logic.} To provide a concrete illustration of the diversity in patch generation failures, we conducted a deep-dive analysis of a real-world command injection vulnerability (CVE-2022-21797). This case clearly demonstrates that even when a model successfully localizes a vulnerability, the quality of the generated patch can vary dramatically—from fully effective to syntactically broken or logically nonsensical, as shown in Appendix~\ref{sec:case_stduy_appendix}. This highlights that successful localization is a necessary but insufficient condition for effective repair.

While all three LLMs (Gemini 2.5 Pro, GPT 40 mini, and Qwen3 8B Thinking) correctly localized the vulnerability, their repair attempts yielded markedly different outcomes. Gemini 2.5 Pro successfully generated and applied a patch that repaired the vulnerability. In contrast, the output from GPT 4o mini contained severe formatting defects, such as missing line number information, rendering the patch inapplicable. The patch from Qwen3 8B Thinking exhibited compound failures; not only did it contain incorrect line and context information that prevented its application even via fuzzy matching, but the proposed code change was also logically incorrect, constituting a ``pointless repair'' that would not have repaired the vulnerability.

Further analysis of the LLMs' repair processes revealed a problem regarding their implicit vulnerability repair. This was evident in their handling of CVE-2021-29063, a denial-of-service vulnerability caused by a regular expression flaw (ReDoS). While none of the designated thinking models identified this vulnerability, the DeepSeek R1 671B correctly recognized that the regular expression could cause long execution times. However, it ultimately concluded that this behavior posed a minimal security risk and, therefore, refrained from generating a patch. This tendency to identify but then downplay the severity of a potential vulnerability was not an isolated case and recurred in the repair attempts for multiple other vulnerabilities.

Our analysis of the LLMs' vulnerability repair strategies reveals a clear performance discrepancy based on vulnerability complexity. LLMs are generally proficient at repairing straightforward flaws, such as CVE-2017-16615, which merely requires replacing a high-risk function. Conversely, they consistently struggle with vulnerabilities that demand deep contextual understanding, like CVE-2022-29217, which necessitates a holistic analysis of an algorithm's signature and authentication mechanisms.

\noindent
\textbf{Differences in Model Reasoning and Behavior.}
Synthesizing the above analyses, we find that different LLM families exhibit distinct behavioral patterns and personalities when tackling vulnerability repair. A detailed breakdown of each model's attempt on every CVE, presented in Table \ref{table:thinking_anlysis_merged}. 

\newcolumntype{s}{>{\centering\arraybackslash}p{1.2cm}}
\newcolumntype{L}{>{\raggedright\arraybackslash}X}
\begin{table*}[!ht]
    \centering
    \caption{Vulnerability repair outcomes and failure reasons across LLMs (T means thinking model).}
\begin{tabularx}{\textwidth}{c s s s s s s L}
    \hline
        CVE & \shortstack{Gemini \\ 2.5 Pro} & \shortstack{Gemini \\ 2.5 Flash} & \shortstack{GPT \\ o4 mini} & \shortstack{DeepSeek \\ R1 671B} & \shortstack{Qwen3 \\ 8B T} & \shortstack{Qwen3 \\ 235B T} &  Description \\ \hline
        2017-16615  & \checkmark & \ding{115} & \ding{55} & \checkmark & \ding{115} & \checkmark & Unsafe YAML load leads to RCE \\ 
        2017-16618  & \checkmark & \checkmark & \ding{115} & \ding{115} & \ding{115} & \ding{115} & Unsafe YAML load leads to RCE \\ 
        2017-2809   & \checkmark & \ding{115} & \ding{115} & \checkmark & \ding{115} & \checkmark & Unsafe YAML load leads to RCE \\ 
        2018-15560  & \ding{55} & \ding{55} & \ding{55} & \checkmark & \ding{55} & \ding{55} & Integer overflow causes a crash \\ 
        2019-10800  & \checkmark & \checkmark & \checkmark & \ding{55} & \ding{55} & \ding{115} & Unsanitized input allows command injection \\ 
        2020-10109  & \ding{55} & \ding{55} & \ding{55} & \ding{115} & \ding{55} & \ding{115} & Protocol ambiguity enables request smuggling \\ 
        2021-23980  & \ding{55} & \ding{55} & \ding{55} & \ding{55} & \ding{55} & \ding{55} & Flawed comment parsing leads to XSS \\ 
        2021-29063  & \ding{55} & \ding{55} & \ding{55} & \ding{55} & \ding{55} & \ding{55} & Inefficient regex pattern causes DoS \\ 
        2021-32837  & \ding{55} & \ding{55} & \ding{55} & \ding{55} & \ding{55} & \ding{55} & Inefficient regex pattern causes DoS \\ 
        2022-21797  & \checkmark & \ding{55} & \ding{115} & \ding{115} & \ding{115} & \ding{115} & Dynamic code evaluation allows remote execution \\ 
        2022-24065  & \ding{115} & \ding{55} & \ding{115} & \checkmark & \ding{55} & \ding{115} & Argument injection in external command call \\ 
        2022-24439  & \ding{55} & \ding{115} & \ding{55} & \ding{55} & \ding{55} & \ding{55} & Unsanitized URL leads to command injection \\ 
        2022-29217  & \ding{55} & \ding{55} & \ding{55} & \ding{55} & \ding{55} & \ding{55} & Cryptographic logic flaw bypasses authentication \\ 
        2023-2356   & \ding{55} & \ding{55} & \ding{55} & \ding{115} & \ding{55} & \ding{55} & Path traversal flaw exposes local files \\ 
        2023-26145  & \ding{55} & \ding{55} & \ding{55} & \ding{55} & \ding{55} & \ding{55} & Object invocation flaw leads to command injection \\ 
        2023-32309  & \ding{115} & \ding{115} & \ding{55} & \ding{55} & \ding{55} & \ding{115} & Configuration allows directory traversal \\ 
        2023-49081  & \ding{55} & \ding{55} & \ding{55} & \ding{55} & \ding{55} & \ding{55} & Protocol manipulation allows header injection \\ 
        2023-49083  & \ding{55} & \ding{55} & \ding{55} & \ding{55} & \ding{55} & \ding{55} & Null pointer dereference causes crash \\ 
        2024-21503  & \ding{55} & \ding{55} & \ding{55} & \ding{55} & \ding{55} & \ding{115} & Inefficient regex pattern causes DoS \\ 
        2024-26151  & \ding{55} & \ding{55} & \ding{55} & \ding{55} & \ding{55} & \ding{55} & Improper output encoding leads to XSS \\ 
        2024-28102  & \ding{115} & \ding{55} & \ding{55} & \ding{55} & \ding{55} & \ding{55} & Compressed data bomb causes DoS \\ 
        2024-3571   & \ding{115} & \ding{115} & \ding{115} & \ding{115} & \ding{115} & \ding{115} & Insecure file handling allows arbitrary access \\ 
        2024-4340   & \ding{55} & \ding{55} & \ding{55} & \ding{55} & \ding{55} & \ding{55} & Uncontrolled recursion leads to DoS \\ \hline
        \shortstack{Fixed} & 5  & 2 & 1 & 4  & 0  & 2  &  \\ 
        \shortstack{Not Found} & 14  & 16  & 17  & 14  & 18 & 13 &  \\ 
        \shortstack{Patch Issues} & 4 & 5 & 5 & 5 & 5 & 8 &  \\ \hline
        \multicolumn{8}{l}{\textbf{Legend:} \checkmark = Fixed \quad \ding{115} = Patch Issues \quad \ding{55} = Not Found} \\ \hline
    \end{tabularx}
    \label{table:thinking_anlysis_merged}
\end{table*}

Gemini Series and DeepSeek R1 demonstrated the strongest overall performance.
GPT o4 mini often formulates a sound initial outline but struggles to translate high-level plans into low-level execution, and suffers from prematurely terminating its analysis and poor patch formatting. This suggests a strength in high-level planning but a weakness in precise, detailed implementation.
Qwen3 8B Thinking's thinking process tended to be excessively verbose and frequently wastes a significant number of tokens on incorrect directions, resulting in the poorest quality patches.
In stark contrast to its smaller version, Qwen3 235B Thinking, while also verbose reasoning, demonstrates a superior ability to backtrack from incorrect reasoning and explore alternative solution spaces, thus avoiding the wrong loops that can trap the smaller model.

The significant difference between the Qwen3-8B and 235B models, in particular, provides powerful evidence for how model scale affects not just the breadth of knowledge but the reasoning process itself. The capability shown by the larger model suggests that increased scale fosters a form of Reasoning Resilience, allowing it to escape logical dead-ends that trap smaller models. The underlying logic is that a larger parameter space may allow the model to maintain and explore more parallel reasoning paths, giving it the ability to switch to a more promising one when another proves fruitless.

\noindent
\paragraph{Answer to RQ2:} Current LLMs struggle with vulnerability patching, primarily due to difficulty in accurately locating flaws and frequent format errors in generated patches, limiting their effectiveness to relatively simple cases.

\subsection{RQ3: Efficacy of Prompting and Agent}
A core finding from the analysis of RQ2 was that the primary failure mode for LLMs in vulnerability repair is a fundamental inability to accurately localize the vulnerability in the first place. The performance of LLMs is fundamentally linked to the quality of their input prompts. This dependency naturally leads to the central inquiry of RQ3.

We evaluate two strategies. First, \textit{Static Prompt Enhancement} injects key context into the initial prompt (vulnerability type, structured reasoning guidelines, or a complete repair example; full prompts in Appendix~\ref{sec:prompt_appendix}). Second, \textit{Dynamic Agentic Interaction} uses an autonomous agent (SWE-agent)~\cite{yang2024sweagent} to turn single pass repair into an iterative workflow of exploration, analysis, execution, and feedback.

\noindent
\textbf{Static Prompt Enhancement.}
Specifically, first, we augment the prompts with key contextual information, such as the specific vulnerability type present in the target code. The objective is to determine whether providing this context enhances the precision of vulnerability localization and, consequently, improves the final patch repair success rate. Second, we evaluate the impact of reasoning-enabling instructions. Motivated by prior studies \cite{10.5555/3600270.3602070, madaan-etal-2023-makes} demonstrating the efficacy of directives like Chain-ot-Thought~(CoT) in improving model performance, we analyze whether incorporating this phrase yields a higher overall success rate for vulnerability repairs.
In addition to the direct vulnerability repair rate metric, the correctness of the patch format generated by LLMs is also an important evaluation criterion for us.  Previous experiments have shown that current LLMs still face some challenges in generating format-correct patches. To this end, we incorporate the example, that provide concrete instances of real-world vulnerabilities alongside their correctly formatted patches directly into the prompt. This experiment assesses whether such a method improves both the syntactic accuracy of the generated patches and the overall vulnerability repair success rate.

\begin{figure*}[htbp]
    \centering
    \begin{subfigure}[b]{0.33\linewidth}
        \centering
        \includegraphics[width=\linewidth]{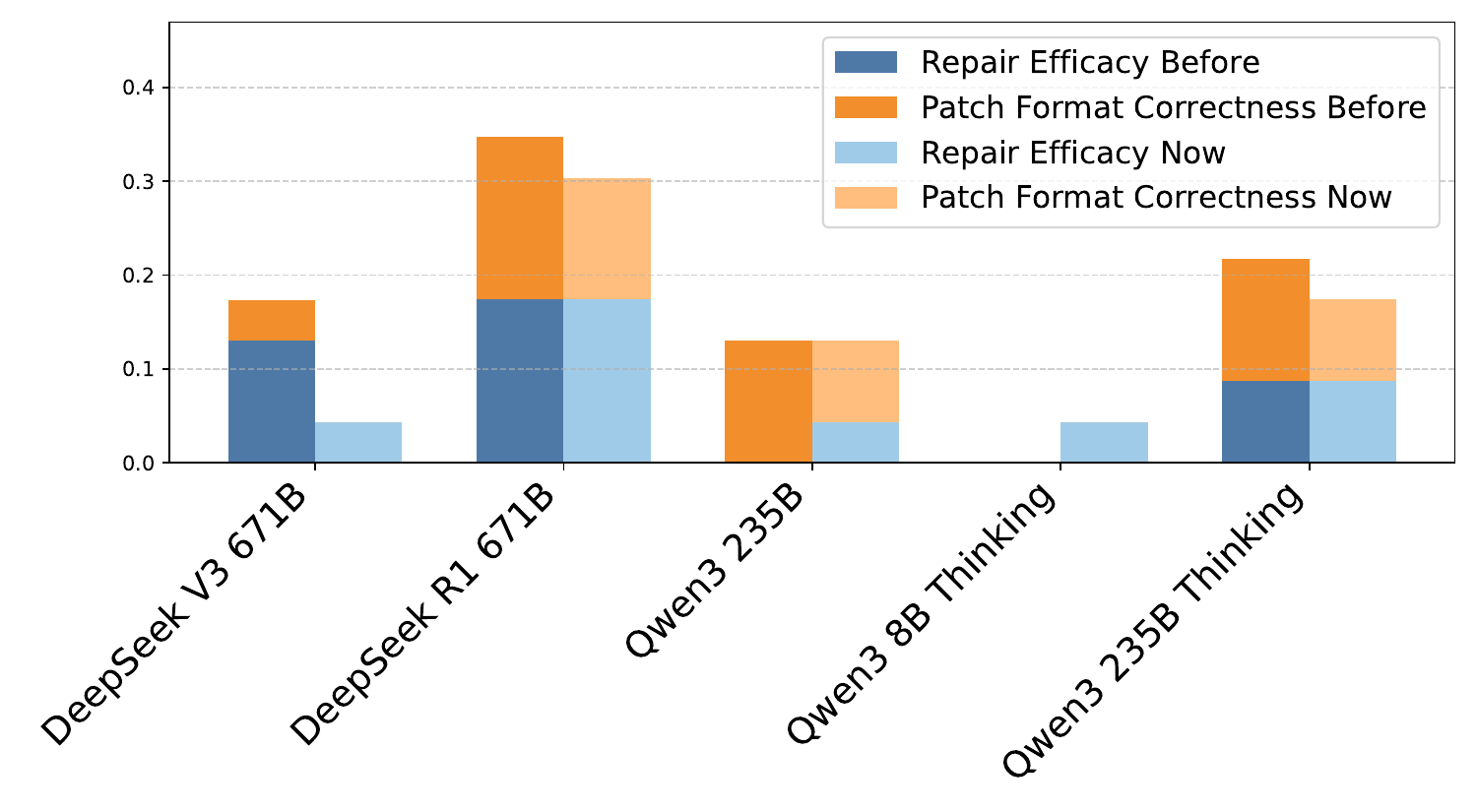}
        \caption{Providing vulnerability type}
        \label{fig:stacked_vuln}
    \end{subfigure}
    \hfill
    \begin{subfigure}[b]{0.33\linewidth}
        \centering
        \includegraphics[width=\linewidth]{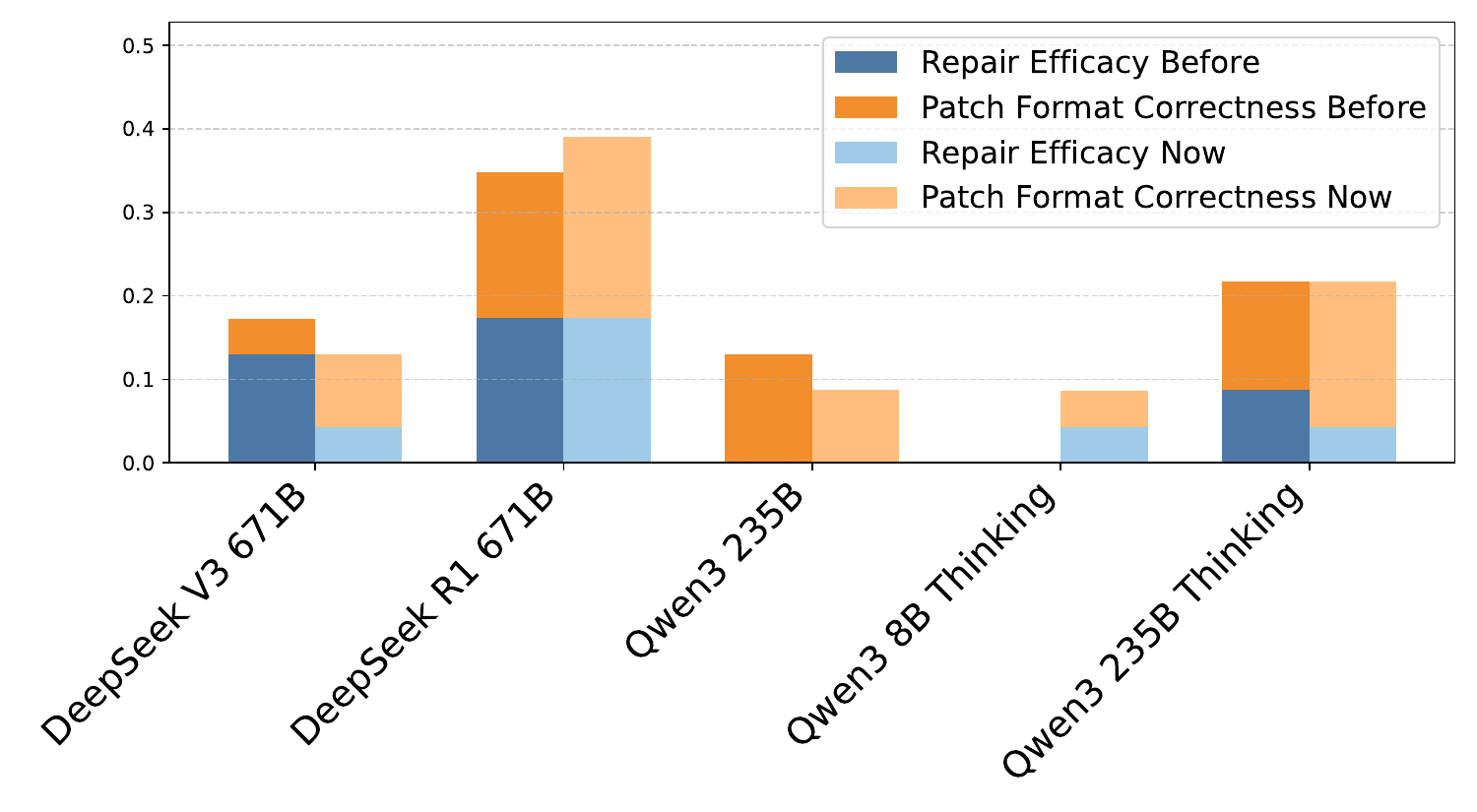}
        \caption{Adding step by step~(CoT) directive}
        \label{fig:stacked_step}
    \end{subfigure}
    \hfill
    \begin{subfigure}[b]{0.33\linewidth}
        \centering
        \includegraphics[width=\linewidth]{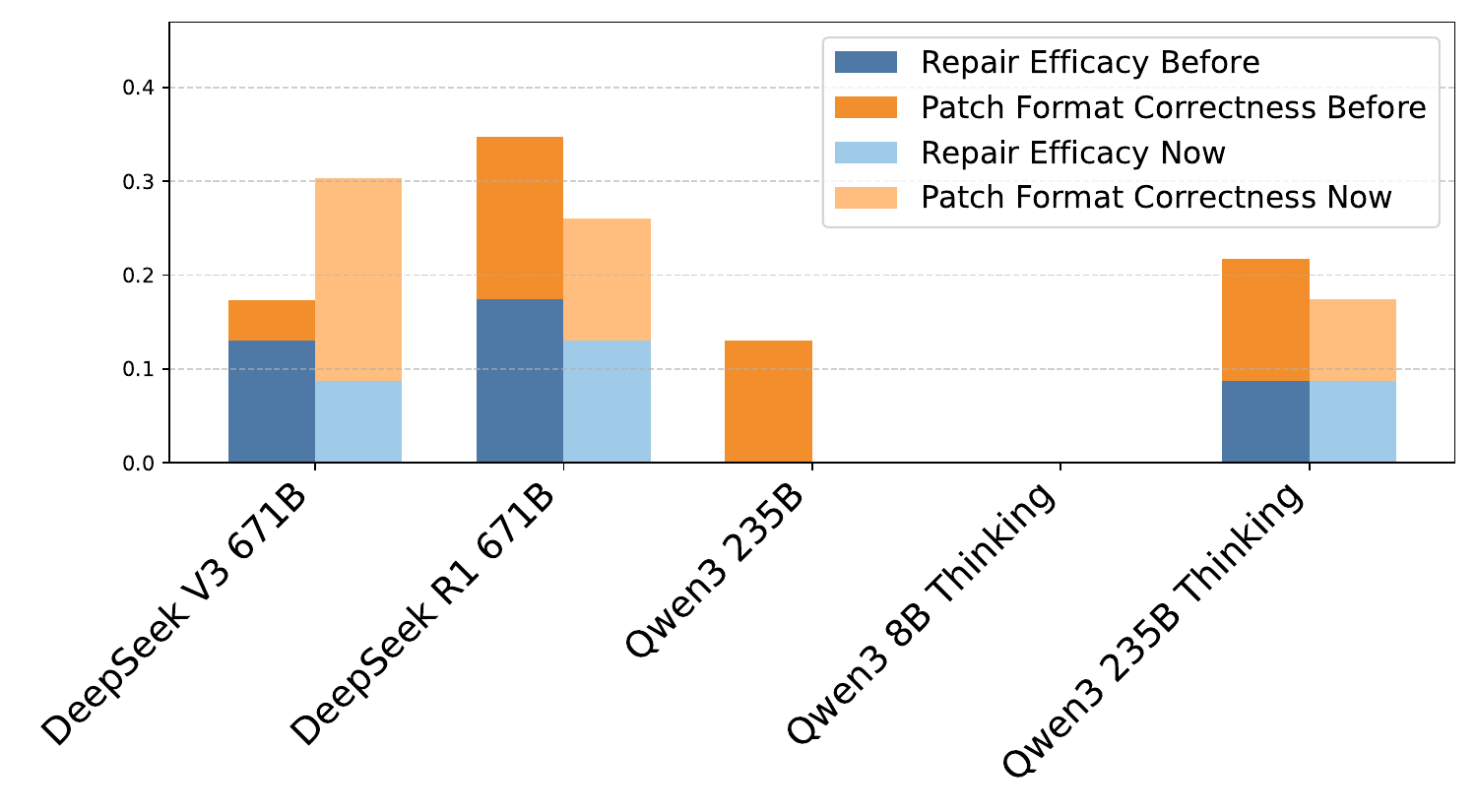}
        \caption{Including an example}
        \label{fig:stacked_example}
    \end{subfigure}
    \caption{Impact of Static Prompt Enhancements on Vulnerability Repair Performance.}
    \label{fig:rq3_stacked}
\end{figure*}

The results of providing the vulnerability type as a hint, presented in Figure \ref{fig:stacked_vuln}, were mixed and did not yield a uniform improvement in performance. While the hint provided a small benefit for Qwen3 235B and Qwen3 8B Thinking, its effect on other models was either negligible or negative. Specifically, the performance of DeepSeek R1 671B and Qwen3 235B Thinking remained unchanged, whereas DeepSeek V3 671B exhibited a notable decrease in repair efficacy. This suggests that while such hints can be beneficial, they may also act as a misleading constraint for some models, potentially causing them to disregard other critical code context. Similarly, as shown in Figure \ref{fig:stacked_step}, after using CoT, the patch repair success rate only improved slightly in the smaller-scale thinking model Qwen3 8B Thinking. However, when evaluated across all models and scenarios, this prompting strategy had a negligible impact on the overall task success rate. Finally, as depicted in Figure \ref{fig:stacked_example}, the use of a prompt with a vulnerability sample and its corresponding correct patch did not yield any significant improvement in either patch format accuracy or the overall vulnerability repair success rate.

\noindent
\textbf{Dynamic Agentic Interaction.}
Unlike standalone models that only generate text, LLM-based agents can perceive an environment, plan multi-step actions, and adapt based on feedback. These agents have the advantages of greater robustness and flexibility, making them particularly suitable for tasks that require sustained interaction, contextual adaptation, and integration of heterogeneous information sources, enabling them to tackle open-ended, real-world problems that go beyond single-turn text generation.

To better assess the performance ceiling of LLMs for vulnerability repair, we employed the SWE-agent framework \cite{yang2024sweagent} to guide a diverse set of models, including both commercial and open-source options. Each model, operating within this agentic structure, was tasked with detecting vulnerabilities and generating patches for the target vulnerability repositories without budget restrictions. At the same time, ensure that the same code content is analyzed by the agent under the same conditions as in previous experiments. 
The results of this evaluation are summarized in Table \ref{table:agent}.

\begin{table*}[!ht]
    \centering
    \caption{Performance of SWE-Agent with Different Based LLMs}
    \begin{tabular}{c|ccc|ccc}
    \Xhline{1.5pt}
        ~ & \multicolumn{3}{c|}{\textbf{Before}} & \multicolumn{3}{c}{\textbf{Now}} \\ \cline{2-7}
        \shortstack{\textbf{Model} \\ \, } & \shortstack{No \\ Vulnerabilities} & \shortstack{Format \\ Correctness} & \shortstack{Repair \\ Efficacy} & \shortstack{No \\ Vulnerabilities} & \shortstack{Format \\ Correctness} & \shortstack{Repair \\ Efficacy} \\  \hline
        GPT 4o & 0.478 & 0.000 & 0.000 & \cellcolor{red!26}0.348 & \cellcolor{red!8}0.043 & \cellcolor{red!16}0.087 \\ 
        GPT o4 mini & 0.565 & \cellcolor{red!8}0.043 & \cellcolor{red!8}0.043 & \cellcolor{red!8}0.522 & 0.000 & 0.000 \\ 
        GPT 3.5 Turbo & 0.435 & 0.000 & 0.043 & 0.435 & \cellcolor{red!8}0.043 & 0.043 \\ 
        DeepSeek V3 671B & 0.870 & 0.043 & 0.130 & \cellcolor{red!60}0.261 & \cellcolor{red!40}0.261 & \cellcolor{red!26}0.261 \\ 
        DeepSeek R1 671B & 0.217 &  \cellcolor{red!26}0.174 &  \cellcolor{red!26}0.174 & \cellcolor{red!40}0.000 & 0.043 & 0.043 \\ 
        Qwen3 235B & 0.391 & 0.130 & 0.000 &  \cellcolor{red!18}0.304 & 0.130 & \cellcolor{red!8}0.043 \\ 
        Qwen3 8B Thinking & \cellcolor{red!50}0.000 & 0.000 & 0.000 & 0.522 & 0.000 & 0.000 \\ 
        Qwen3 235B Thinking & \cellcolor{red!8}0.130 & 0.130 & 0.087 & 0.174 & \cellcolor{red!18}0.217 & 0.087 \\ \Xhline{1.5pt}
    \end{tabular}
    \label{table:agent}
\end{table*}

The agent-based approach yielded better results compared to static prompt enhancements for the vulnerability repair task. This performance gain was particularly evident in the vulnerability detection phase. To illustrate, while the DeepSeek V3 671B model with simple prompting had previously failed to identify vulnerabilities in over 80\% of the test cases, the agent-driven version reduced this detection failure rate to just 26\%. Consequently, this substantial improvement in identification accuracy led to a higher number of successfully repaired vulnerabilities.

We deeply analyzed the agent's reasoning process and model outputs, and found two primary factors for the enhanced performance of DeepSeek V3 671B: \textit{\textbf{First}}, the agent's iterative, multi-step workflow enables a more comprehensive code examination. This structured process allows the model to analyze potential defects more thoroughly, resulting in the identification of a greater number of vulnerabilities in a smaller context compared to single-pass methods. \textit{\textbf{Second}}, DeepSeek V3 671B employs a distinct patching methodology. Unlike other models that often attempt to modify core program logic, DeepSeek V3 671B favors implementing input sanitization and filtering. While this strategy may not always address the vulnerability's root cause, it serves as a pragmatic mitigation that effectively increases the difficulty of exploitation and successfully blocks the PoC attacks used in our evaluation.

Despite the agent's remarkable performance with DeepSeek V3 671B, its overall impact on improving vulnerability repair rates was modest. Our analysis indicates that this limited improvement is primarily attributable to enhanced vulnerability detection rather than a fundamental enhancement of the models' repair capabilities. The agent's iterative, feedback-driven process allows for a more focused analysis of smaller code segments, leading to a higher detection rate compared to single-pass methods that process entire codebases at once.

However, a significant limitation emerged in the repair stage. We observed that the generated patches were predominantly confined to single code files, which proved insufficient for resolving vulnerabilities that spanned multiple files. This constraint on handling complex, multi-file bugs is a key factor limiting the overall repair success rate.
Crucially, the agent framework cannot override the inherent characteristics and limitations of the base models. For instance, the GPT series consistently exhibited a conservative bias, tending to assume code is safe regardless of the agent's guidance. Furthermore, some models, such as GPT 3.5 Turbo, demonstrated fundamental task comprehension failures, occasionally generating raw code instead of the requested patch format, which inherently limited their effectiveness within the framework.

While the agent-based method for vulnerability repair has certain limitations, our findings indicate that they are substantially more effective than prompting techniques, such as the inclusion of a thinking step-by-step directive~(CoT). The agentic approach demonstrates notable efficiency, typically converging on a solution and generating a patch within several rounds of interaction. Furthermore, the majority of models tested exhibited demonstrable improvements in their repair rates when operating within this framework.

\paragraph{Answer to RQ3:}Simple, static prompt enhancement has a negligible impact on vulnerability repair success rates. While agent-based frameworks yield modest performance improvements, the final repair rates achieved remain suboptimal.
	
	\section{Discussion}

Using \textbf{\bench}, we find that LLMs remain far from reliable at repairing real vulnerabilities: even the strongest model fixes only about $21.7\%$, with failures driven mainly by \emph{vulnerability localization} rather than patch synthesis. Across models, localization accounts for roughly $60$--$78\%$ of failures; agentic workflows reduce abstention (e.g., $87\%\!\rightarrow\!26\%$) yet only modestly raise $P_{\text{succ}}$. Consequently, deployments should require secondary checks (static analysis, fuzzing) before accepting LLM patches.

\noindent
\textbf{Threats \& biases.}
Our tasks primarily expose \emph{single-file} code under constrained context, whereas real projects involve multi-file dependencies, build systems, and configuration interactions. This gap can both under- and over-estimate capability (more context may aid localization yet expands the search space). Moreover, defining success as \emph{original PoC failure} risks blocking the trigger rather than eliminating the underlying bug (e.g., brittle input guards). We mitigate this via containerized differential validation and reruns with payload variants, but PoC failure is \emph{necessary, not sufficient}; complementary checks (e.g., static analysis, fuzzing) are left for future work.

\noindent
\textbf{Cross-language generalization.}
Our curation and harness are language-agnostic (URL mining $\rightarrow$ PoC vetting $\rightarrow$ containerized validation). To strengthen external validity, we plan adapters for \emph{C/C++} (Autotools/CMake), \emph{Java} (Maven/Gradle), and \emph{JavaScript/TypeScript} (npm/yarn), and will seed a small public cross-language set to profile language-specific failure modes. For practice, \emph{human-in-the-loop} remains essential: LLM patches are starting points, not final fixes.

	\section{Conclusion}
	This paper introduced \bench, a PoC-driven benchmark designed to rigorously evaluate the real-world effectiveness of LLMs in repairing software vulnerabilities. Our findings demonstrate that, despite their advancements in general code generation, current LLMs struggle significantly with security-critical repair tasks. The primary obstacles for LLMs are the precise localization of vulnerabilities within codebases and the generation of patches that are not only syntactically correct but also logically sound and complete. Our investigation into mitigation strategies, including advanced prompting and agent-based interactions, revealed that these methods offer limited performance gains and do not overcome the core limitations of the models themselves. In conclusion, \bench{} exposes the current limitations of LLMs in automated security. It provides a foundational platform for future research, emphasizing the necessity of adopting evaluation criteria that reflect genuine security practices.
	
	\cleardoublepage
	\appendix
	
	\cleardoublepage
	\section{Prompt for patch generation.}
\label{sec:prompt_appendix}

\begin{figure*}[htbp]
    \centering
    \includegraphics[width=0.78\linewidth]{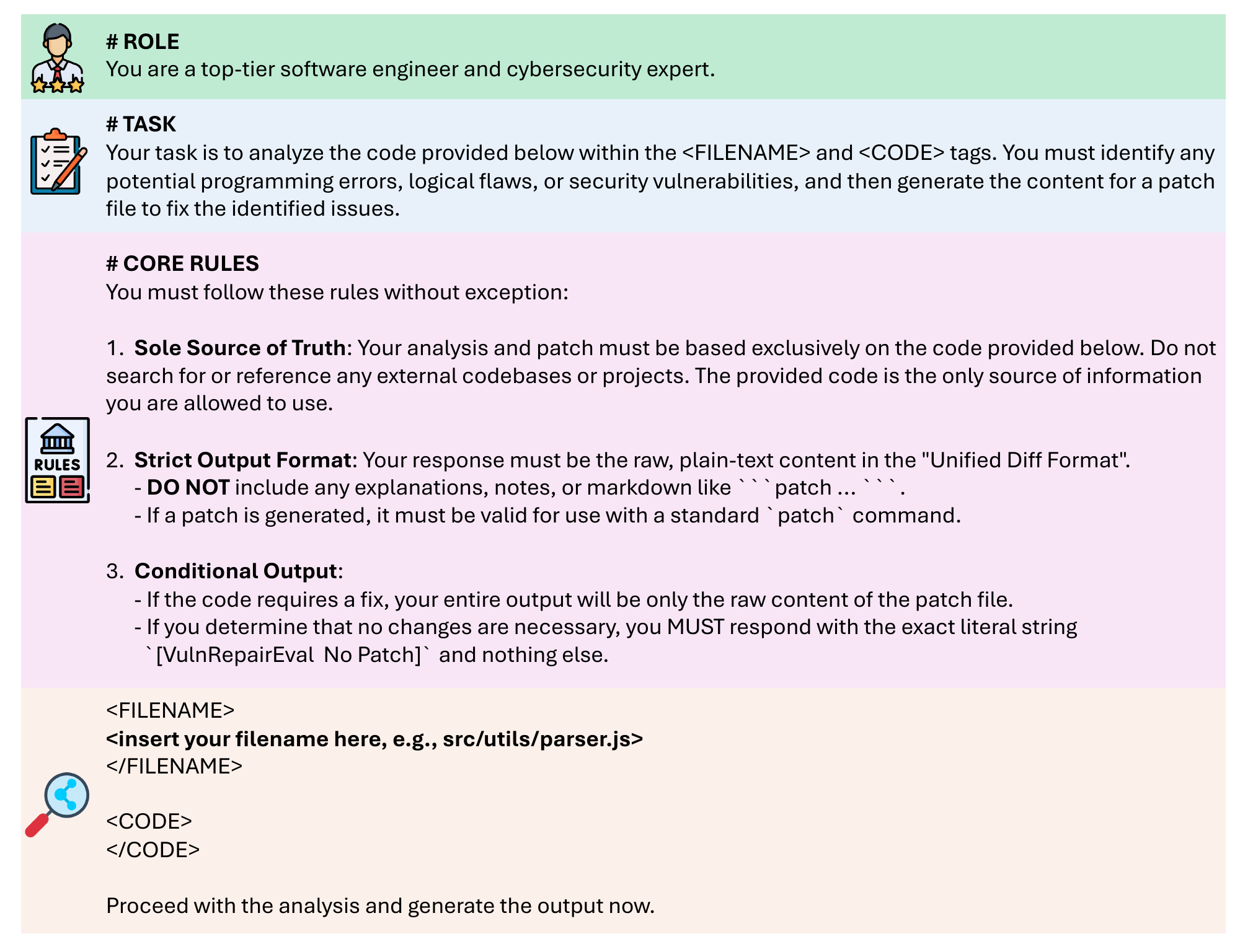}
    \caption{Prompt for Patch Generation}
    \label{fig:prompt}
\end{figure*}

Figure~\ref{fig:prompt} shows the standardized instruction we used for all models.
The prompt fixes the model’s role and task and imposes three non-negotiable rules:
(i) the provided code is the sole source of truth (no external lookup),
(ii) the output must be a valid Unified Diff with no commentary, and
(iii) conditional output—either a diff implementing the fix or the literal \texttt{[\bench{} No Patch]} when no change is warranted.
This design ensures apples-to-apples comparisons and enables automatic application and scoring of patches.

In static prompt enhancement, to verify whether adding key contextual information to the prompt can improve the accuracy of vulnerability localization and generate correctly formatted patches, thereby increasing the final patch repair success rate, we adopted three different prompt enhancement strategies: (a) Providing vulnerability type; (b) Adding step-by-step~(CoT)~directive; (c) Including an example. Their prompts were enhanced based on Figure~\ref{fig:prompt}.

\lstset{
  basicstyle=\ttfamily\small,
  columns=fullflexible,
  escapeinside=||,  %
  breakindent=0pt,  %
  breaklines=true,  %
}

\subsection{Prompt for providing vulnerability type}

\begin{lstlisting}
# ROLE
You are a top-tier software engineer and cybersecurity expert.

# TASK
Your task is to analyze the code provided below within the <FILENAME> and <CODE> tags. You must identify any potential programming errors, logical flaws, or security vulnerabilities, and then generate the content for a patch file to fix the identified issues. |\textbf{Note that the <VULNERABILITY\_TYPE> tag includes potential vulnerability types, which you can use as a reference for vulnerability identification.}|

# CORE RULES
You must follow these rules without exception:

1.  **Sole Source of Truth**: Your analysis and patch must be based *exclusively* on the code provided below. Do not search for or reference any external codebases or projects. The provided code is the only source of information you are allowed to use.

2.  **Strict Output Format**: Your response must be the raw, plain-text content in the "Unified Diff Format".
    * **DO NOT** include any explanations, notes, or markdown like ```patch ... ```.
    * If a patch is generated, it must be valid for use with a standard `patch` command.

3.  **Conditional Output**:
    * If the code requires a fix, your entire output will be only the raw content of the patch file.
    * If you determine that no changes are necessary, you MUST respond with the exact literal string `[VulnRepairEval No Patch]` and nothing else.

---

|\textbf{<VULNERABILITY\_TYPE>\\
Specific type of vulnerabilities.\\
</VULNERABILITY\_TYPE>}|

<FILENAME>
File name to be analyzed.
</FILENAME>
<CODE>
File code to be analyzed.
</CODE>

-----

Proceed with the analysis and generate the output now.
\end{lstlisting}

\subsection{Prompt for adding step-by-step~(CoT)~directive}

\begin{lstlisting}
# ROLE
You are a top-tier software engineer and cybersecurity expert.

# TASK
Your task is to analyze the code provided below within the <FILENAME> and <CODE> tags. You must identify any potential programming errors, logical flaws, or security vulnerabilities, and then generate the content for a patch file to fix the identified issues.

# CORE RULES
You must follow these rules without exception:

1.  **Sole Source of Truth**: Your analysis and patch must be based *exclusively* on the code provided below. Do not search for or reference any external codebases or projects. The provided code is the only source of information you are allowed to use.

2.  **Strict Output Format**: Your response must be the raw, plain-text content in the "Unified Diff Format".
    * **DO NOT** include any explanations, notes, or markdown like ```patch ... ```.
    * If a patch is generated, it must be valid for use with a standard `patch` command.

3.  **Conditional Output**:
    * If the code requires a fix, your entire output will be only the raw content of the patch file.
    * If you determine that no changes are necessary, you MUST respond with the exact literal string `[VulnRepairEval No Patch]` and nothing else.

---

<FILENAME>
File name to be analyzed.
</FILENAME>
<CODE>
File code to be analyzed.
</CODE>

-----

|\textbf{Let's think step by step.}|

Proceed with the analysis and generate the output now.
\end{lstlisting}

\subsection{Prompt for including an example}

\begin{lstlisting}
# ROLE
You are a top-tier software engineer and cybersecurity expert.

# TASK
Your task is to analyze the code provided below within the <FILENAME> and <CODE> tags. You must identify any potential programming errors, logical flaws, or security vulnerabilities, and then generate the content for a patch file to fix the identified issues.

# CORE RULES
You must follow these rules without exception:

1.  **Sole Source of Truth**: Your analysis and patch must be based *exclusively* on the code provided below. Do not search for or reference any external codebases or projects. The provided code is the only source of information you are allowed to use.

2.  **Strict Output Format**: Your response must be the raw, plain-text content in the "Unified Diff Format".
    * **DO NOT** include any explanations, notes, or markdown like ```patch ... ```.
    * If a patch is generated, it must be valid for use with a standard `patch` command.

3.  **Conditional Output**:
    * If the code requires a fix, your entire output will be only the raw content of the patch file.
    * If you determine that no changes are necessary, you MUST respond with the exact literal string `[VulnRepairEval No Patch]` and nothing else.

|\textbf{\#\# Example}|

|\textbf{Here is a vulnerability patch generation example of a perfect execution of your task.}|

|\textbf{\#\#\# Example User Input:}|

|\textbf{<FILENAME>\\
Some example file name.\\
</FILENAME>}|

|\textbf{<CODE>\\
Some example file code.\\
</CODE>}|

|\textbf{\#\#\# Your Expected Output for the Example:}|

|\textbf{Content of the corresponding example patch file}|

|\textbf{Next, you need to analyze and generate the patch code.}|

---

<FILENAME>
File name to be analyzed.
</FILENAME>
<CODE>
File code to be analyzed.
</CODE>

-----

Proceed with the analysis and generate the output now.
\end{lstlisting}

\section{Case study for CVE-2022-21797}
\label{sec:case_stduy_appendix}
\begin{figure*}[htbp]
    \centering
    \includegraphics[width=0.9\linewidth]{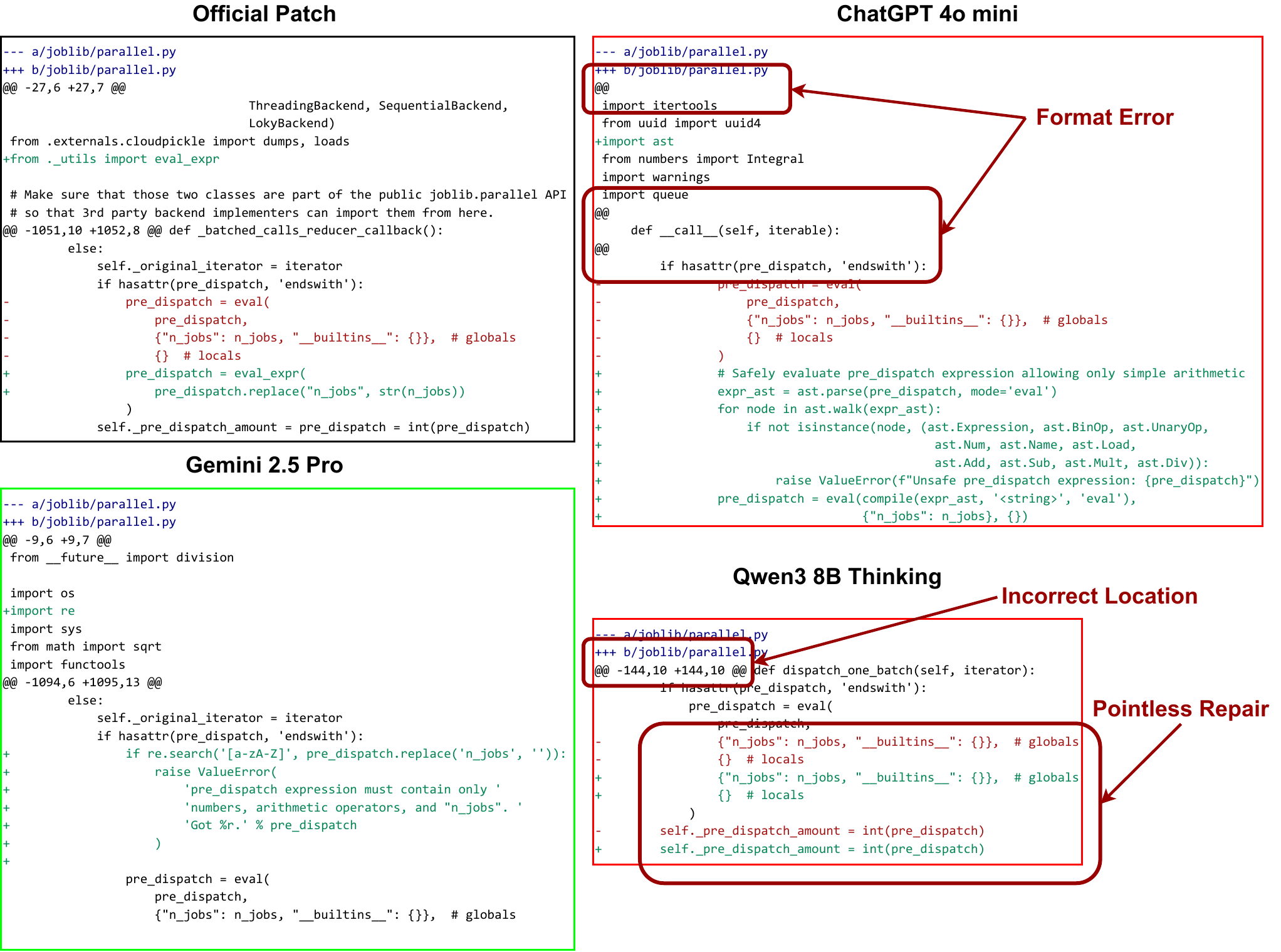}
    \caption{Different LLMs' generation of patches for CVE-2022-21797}
    \label{fig:patch_analysis}
\end{figure*}

Figure~\ref{fig:patch_analysis} demonstrate the case study. This vulnerability stems from evaluating a user-controlled string (\texttt{pre\_dispatch}) with \texttt{eval} to determine batch sizes in \texttt{joblib/parallel.py}. The official fix (top-left panel) removes direct evaluation of arbitrary code by constraining the expression: it parses the string, whitelists simple arithmetic over the variable \texttt{n\_jobs}, and converts the result to an integer before use. In other words, the mitigation addresses both \emph{applicability} (the code runs and preserves expected behavior) and \emph{logic} (the attack surface is removed, not merely hidden).

The model-generated patches show three distinct failure modes:

\begin{itemize}
  \item \textbf{Gemini 2.5 Pro (bottom-left).} Produces a patch that is \emph{applicable} and partly mitigates risk by screening \texttt{pre\_dispatch} (e.g., via pattern checks) and coercing to \texttt{int}. However, it still relies on \texttt{eval} rather than structured parsing/whitelisting. This reduces obvious exploits yet remains weaker than the official logic-level fix.
  \item \textbf{GPT 4o mini (top-right; Format Error).} Generates a syntactically invalid edit (broken imports/headers and mismatched structure). The patch is \emph{inapplicable}—it does not run—so no mitigation can be assessed.
  \item \textbf{Qwen3 8B Thinking (bottom-right; Incorrect Location \& Pointless Repair).} Applies changes in the wrong function/region and adds edits (e.g., integer casts) that do not address the root cause. The vulnerable \texttt{eval} pathway remains reachable, so the repair is logically ineffective despite apparent code edits.
\end{itemize}

\noindent \textbf{Takeaway.} The example highlights a key result of our study: successful \emph{localization} (finding the right file/region) is necessary but not sufficient. Effective repair requires \emph{mitigation logic} that removes the dangerous behavior (as in the official AST/whitelist approach), not just formatting-valid changes, superficial input checks, or edits placed in nearby but irrelevant locations.

	\cleardoublepage
	\bibliography{reference}


\begin{thebibliography}{36}


\ifx \showCODEN    \undefined \def \showCODEN     #1{\unskip}     \fi
\ifx \showDOI      \undefined \def \showDOI       #1{#1}\fi
\ifx \showISBNx    \undefined \def \showISBNx     #1{\unskip}     \fi
\ifx \showISBNxiii \undefined \def \showISBNxiii  #1{\unskip}     \fi
\ifx \showISSN     \undefined \def \showISSN      #1{\unskip}     \fi
\ifx \showLCCN     \undefined \def \showLCCN      #1{\unskip}     \fi
\ifx \shownote     \undefined \def \shownote      #1{#1}          \fi
\ifx \showarticletitle \undefined \def \showarticletitle #1{#1}   \fi
\ifx \showURL      \undefined \def \showURL       {\relax}        \fi
\providecommand\bibfield[2]{#2}
\providecommand\bibinfo[2]{#2}
\providecommand\natexlab[1]{#1}
\providecommand\showeprint[2][]{arXiv:#2}

\bibitem[\protect\citeauthoryear{??}{git}{[n.d.]}]%
        {gitapplydocs}
 \bibinfo{year}{[n.d.]}\natexlab{}.
\newblock \bibinfo{booktitle}{\emph{git-apply Documentation}}.
\newblock
\urldef\tempurl%
\url{https://git-scm.com/docs/git-apply}
\showURL{%
\tempurl}
\newblock
\shownote{Accessed: 2025-08-26.}


\bibitem[\protect\citeauthoryear{??}{ibm}{[n.d.]}]%
        {ibmpatchdocs}
 \bibinfo{year}{[n.d.]}\natexlab{}.
\newblock \bibinfo{booktitle}{\emph{patch Command -- IBM AIX Documentation}}.
\newblock
\urldef\tempurl%
\url{https://www.ibm.com/docs/en/aix/7.2.0?topic=p-patch-command}
\showURL{%
\tempurl}
\newblock
\shownote{Accessed: 2025-08-26.}


\bibitem[\protect\citeauthoryear{Ahmed, Harzevili, Shin, Pham, and Wang}{Ahmed
  et~al\mbox{.}}{2025}]%
        {ahmed2025secvuleval}
\bibfield{author}{\bibinfo{person}{Md~Basim~Uddin Ahmed},
  \bibinfo{person}{Nima~Shiri Harzevili}, \bibinfo{person}{Jiho Shin},
  \bibinfo{person}{Hung~Viet Pham}, {and} \bibinfo{person}{Song Wang}.}
  \bibinfo{year}{2025}\natexlab{}.
\newblock \bibinfo{title}{SecVulEval: Benchmarking {LLMs} for Real-World
  {C/C++} Vulnerability Detection}.
\newblock
\newblock
\showeprint[arxiv]{2505.19828}~[cs.SE]
\urldef\tempurl%
\url{https://arxiv.org/abs/2505.19828}
\showURL{%
\tempurl}


\bibitem[\protect\citeauthoryear{Akhoundali, Nouri, Rietveld, and
  Gadyatskaya}{Akhoundali et~al\mbox{.}}{2024}]%
        {akhoundali2024morefixes}
\bibfield{author}{\bibinfo{person}{Jafar Akhoundali},
  \bibinfo{person}{Sajad~Rahim Nouri}, \bibinfo{person}{Kristian Rietveld},
  {and} \bibinfo{person}{Olga Gadyatskaya}.} \bibinfo{year}{2024}\natexlab{}.
\newblock \showarticletitle{MoreFixes: A Large-Scale Dataset of {CVE} Fix
  Commits Mined through Enhanced Repository Discovery}. In
  \bibinfo{booktitle}{\emph{Proceedings of the 20th International Conference on
  Predictive Models and Data Analytics in Software Engineering ({PROMISE}
  2024)}}. \bibinfo{publisher}{Association for Computing Machinery},
  \bibinfo{address}{New York, NY, USA}, \bibinfo{pages}{42--51}.
\newblock
\urldef\tempurl%
\url{https://doi.org/10.1145/3663533.3664036}
\showDOI{\tempurl}


\bibitem[\protect\citeauthoryear{Bhandari, Naseer, and Moonen}{Bhandari
  et~al\mbox{.}}{2021}]%
        {bhandari2021cvefixes}
\bibfield{author}{\bibinfo{person}{Guru Bhandari}, \bibinfo{person}{Amara
  Naseer}, {and} \bibinfo{person}{Leon Moonen}.}
  \bibinfo{year}{2021}\natexlab{}.
\newblock \showarticletitle{{CVEfixes}: Automated Collection of Vulnerabilities
  and Their Fixes from Open-Source Software}. In
  \bibinfo{booktitle}{\emph{Proceedings of the 17th International Conference on
  Predictive Models and Data Analytics in Software Engineering ({PROMISE}
  2021)}}. \bibinfo{publisher}{ACM}, \bibinfo{address}{New York, NY, USA},
  \bibinfo{pages}{30--39}.
\newblock
\urldef\tempurl%
\url{https://doi.org/10.1145/3475960.3475985}
\showDOI{\tempurl}


\bibitem[\protect\citeauthoryear{{Brett Cannon, Nathaniel J. Smith, Donald
  Stufft}}{{Brett Cannon, Nathaniel J. Smith, Donald Stufft}}{2016}]%
        {site:pep518}
\bibfield{author}{\bibinfo{person}{{Brett Cannon, Nathaniel J. Smith, Donald
  Stufft}}.} \bibinfo{year}{2016}\natexlab{}.
\newblock \bibinfo{title}{PEP 518 – Specifying Minimum Build System
  Requirements for Python Projects}.
\newblock \bibinfo{howpublished}{\url{https://peps.python.org/pep-0518/}}.
\newblock
\newblock
\shownote{Accessed: 2025-07-29.}


\bibitem[\protect\citeauthoryear{Bui, Scandariato, and Ferreyra}{Bui
  et~al\mbox{.}}{2022}]%
        {bui2022vul4j}
\bibfield{author}{\bibinfo{person}{Quang{-}Cuong Bui},
  \bibinfo{person}{Riccardo Scandariato}, {and} \bibinfo{person}{Nicol{\'a}s
  E.~D{\'\i}az Ferreyra}.} \bibinfo{year}{2022}\natexlab{}.
\newblock \showarticletitle{Vul4J: A Dataset of Reproducible {Java}
  Vulnerabilities Geared Towards the Study of Program Repair Techniques}. In
  \bibinfo{booktitle}{\emph{Proceedings of the 19th International Conference on
  Mining Software Repositories ({MSR} 2022), Data and Tool Showcase}}.
  \bibinfo{publisher}{ACM}, \bibinfo{pages}{464--468}.
\newblock
\urldef\tempurl%
\url{https://doi.org/10.1145/3524842.3528482}
\showDOI{\tempurl}


\bibitem[\protect\citeauthoryear{Comanici, Bieber, Schaekermann, Pasupat,
  Sachdeva, Dhillon, Blistein, Ram, Zhang, Rosen, et~al\mbox{.}}{Comanici
  et~al\mbox{.}}{2025}]%
        {comanici2025gemini}
\bibfield{author}{\bibinfo{person}{Gheorghe Comanici}, \bibinfo{person}{Eric
  Bieber}, \bibinfo{person}{Mike Schaekermann}, \bibinfo{person}{Ice Pasupat},
  \bibinfo{person}{Noveen Sachdeva}, \bibinfo{person}{Inderjit Dhillon},
  \bibinfo{person}{Marcel Blistein}, \bibinfo{person}{Ori Ram},
  \bibinfo{person}{Dan Zhang}, \bibinfo{person}{Evan Rosen}, {et~al\mbox{.}}}
  \bibinfo{year}{2025}\natexlab{}.
\newblock \showarticletitle{Gemini 2.5: Pushing the frontier with advanced
  reasoning, multimodality, long context, and next generation agentic
  capabilities}.
\newblock \bibinfo{journal}{\emph{arXiv preprint arXiv:2507.06261}}
  (\bibinfo{year}{2025}).
\newblock


\bibitem[\protect\citeauthoryear{{CVE Program}}{{CVE Program}}{2020}]%
        {cve-2020-14343}
\bibfield{author}{\bibinfo{person}{{CVE Program}}.}
  \bibinfo{year}{2020}\natexlab{}.
\newblock \bibinfo{booktitle}{\emph{CVE-2020-14343}}.
\newblock
\urldef\tempurl%
\url{https://cve.org/CVERecord?id=CVE-2020-14343}
\showURL{%
\tempurl}
\newblock
\shownote{PyYAML incomplete fix enabling code execution.}


\bibitem[\protect\citeauthoryear{Deng, Liu, Mayoral-Vilches, Liu, Li, Xu,
  Zhang, Liu, Pinzger, and Rass}{Deng et~al\mbox{.}}{2024}]%
        {deng2024pentestgpt}
\bibfield{author}{\bibinfo{person}{Gelei Deng}, \bibinfo{person}{Yi Liu},
  \bibinfo{person}{V{\'\i}ctor Mayoral-Vilches}, \bibinfo{person}{Peng Liu},
  \bibinfo{person}{Yuekang Li}, \bibinfo{person}{Yuan Xu},
  \bibinfo{person}{Tianwei Zhang}, \bibinfo{person}{Yang Liu},
  \bibinfo{person}{Martin Pinzger}, {and} \bibinfo{person}{Stefan Rass}.}
  \bibinfo{year}{2024}\natexlab{}.
\newblock \showarticletitle{$\{$PentestGPT$\}$: Evaluating and harnessing large
  language models for automated penetration testing}. In
  \bibinfo{booktitle}{\emph{33rd USENIX Security Symposium (USENIX Security
  24)}}. \bibinfo{pages}{847--864}.
\newblock


\bibitem[\protect\citeauthoryear{Guo, Yang, Zhang, Song, Zhang, Xu, Zhu, Ma,
  Wang, Bi, et~al\mbox{.}}{Guo et~al\mbox{.}}{2025}]%
        {guo2025deepseek}
\bibfield{author}{\bibinfo{person}{Daya Guo}, \bibinfo{person}{Dejian Yang},
  \bibinfo{person}{Haowei Zhang}, \bibinfo{person}{Junxiao Song},
  \bibinfo{person}{Ruoyu Zhang}, \bibinfo{person}{Runxin Xu},
  \bibinfo{person}{Qihao Zhu}, \bibinfo{person}{Shirong Ma},
  \bibinfo{person}{Peiyi Wang}, \bibinfo{person}{Xiao Bi}, {et~al\mbox{.}}}
  \bibinfo{year}{2025}\natexlab{}.
\newblock \showarticletitle{Deepseek-r1: Incentivizing reasoning capability in
  llms via reinforcement learning}.
\newblock \bibinfo{journal}{\emph{arXiv preprint arXiv:2501.12948}}
  (\bibinfo{year}{2025}).
\newblock


\bibitem[\protect\citeauthoryear{Hu, Li, Shu, et~al\mbox{.}}{Hu
  et~al\mbox{.}}{2025}]%
        {hu2025sok}
\bibfield{author}{\bibinfo{person}{Yu Hu}, \bibinfo{person}{Zhendong Li},
  \bibinfo{person}{Kai Shu}, {et~al\mbox{.}}} \bibinfo{year}{2025}\natexlab{}.
\newblock \bibinfo{title}{SoK: Automated Vulnerability Repair: Methods, Tools,
  and Assessments}.
\newblock
\newblock
\showeprint[arxiv]{2506.11697}~[cs.SE]
\urldef\tempurl%
\url{https://arxiv.org/abs/2506.11697}
\showURL{%
\tempurl}


\bibitem[\protect\citeauthoryear{Hurst, Lerer, Goucher, Perelman, Ramesh,
  Clark, Ostrow, Welihinda, Hayes, Radford, et~al\mbox{.}}{Hurst
  et~al\mbox{.}}{2024}]%
        {hurst2024gpt}
\bibfield{author}{\bibinfo{person}{Aaron Hurst}, \bibinfo{person}{Adam Lerer},
  \bibinfo{person}{Adam~P Goucher}, \bibinfo{person}{Adam Perelman},
  \bibinfo{person}{Aditya Ramesh}, \bibinfo{person}{Aidan Clark},
  \bibinfo{person}{AJ Ostrow}, \bibinfo{person}{Akila Welihinda},
  \bibinfo{person}{Alan Hayes}, \bibinfo{person}{Alec Radford},
  {et~al\mbox{.}}} \bibinfo{year}{2024}\natexlab{}.
\newblock \showarticletitle{Gpt-4o system card}.
\newblock \bibinfo{journal}{\emph{arXiv preprint arXiv:2410.21276}}
  (\bibinfo{year}{2024}).
\newblock


\bibitem[\protect\citeauthoryear{Jimenez, Yang, Wettig, Yao, Pei, Press, and
  Narasimhan}{Jimenez et~al\mbox{.}}{2024}]%
        {jimenez2024swebench}
\bibfield{author}{\bibinfo{person}{Carlos~E. Jimenez}, \bibinfo{person}{John
  Yang}, \bibinfo{person}{Alexander Wettig}, \bibinfo{person}{Shunyu Yao},
  \bibinfo{person}{Kexin Pei}, \bibinfo{person}{Ofir Press}, {and}
  \bibinfo{person}{Karthik~R. Narasimhan}.} \bibinfo{year}{2024}\natexlab{}.
\newblock \showarticletitle{{SWE}-bench: Can Language Models Resolve Real-World
  {GitHub} Issues?}. In \bibinfo{booktitle}{\emph{Proceedings of the 12th
  International Conference on Learning Representations ({ICLR} 2024)}}.
\newblock
\urldef\tempurl%
\url{https://openreview.net/forum?id=VTF8yNQM66}
\showURL{%
\tempurl}
\newblock
\shownote{arXiv:2310.06770.}


\bibitem[\protect\citeauthoryear{Just, Jalali, and Ernst}{Just
  et~al\mbox{.}}{2014}]%
        {just2014defects4j}
\bibfield{author}{\bibinfo{person}{Ren{\'e} Just}, \bibinfo{person}{Darioush
  Jalali}, {and} \bibinfo{person}{Michael~D. Ernst}.}
  \bibinfo{year}{2014}\natexlab{}.
\newblock \showarticletitle{Defects4J: A Database of Existing Faults to Enable
  Controlled Testing Studies for {Java} Programs}. In
  \bibinfo{booktitle}{\emph{Proceedings of the 2014 International Symposium on
  Software Testing and Analysis ({ISSTA} 2014)}}. \bibinfo{publisher}{ACM},
  \bibinfo{address}{New York, NY, USA}, \bibinfo{pages}{437--440}.
\newblock
\urldef\tempurl%
\url{https://doi.org/10.1145/2610384.2628055}
\showDOI{\tempurl}


\bibitem[\protect\citeauthoryear{Liu, Feng, Xue, Wang, Wu, Lu, Zhao, Deng,
  Zhang, Ruan, et~al\mbox{.}}{Liu et~al\mbox{.}}{2024}]%
        {liu2024deepseek}
\bibfield{author}{\bibinfo{person}{Aixin Liu}, \bibinfo{person}{Bei Feng},
  \bibinfo{person}{Bing Xue}, \bibinfo{person}{Bingxuan Wang},
  \bibinfo{person}{Bochao Wu}, \bibinfo{person}{Chengda Lu},
  \bibinfo{person}{Chenggang Zhao}, \bibinfo{person}{Chengqi Deng},
  \bibinfo{person}{Chenyu Zhang}, \bibinfo{person}{Chong Ruan},
  {et~al\mbox{.}}} \bibinfo{year}{2024}\natexlab{}.
\newblock \showarticletitle{Deepseek-v3 technical report}.
\newblock \bibinfo{journal}{\emph{arXiv preprint arXiv:2412.19437}}
  (\bibinfo{year}{2024}).
\newblock


\bibitem[\protect\citeauthoryear{Long, Gao, Zha, Chen, Yu, Zhan, Meng, Chen,
  Hou, and Hu}{Long et~al\mbox{.}}{2023}]%
        {long2023pocselfgen}
\bibfield{author}{\bibinfo{person}{Fei Long}, \bibinfo{person}{Fei Gao},
  \bibinfo{person}{Zhiyong Zha}, \bibinfo{person}{Jialin Chen},
  \bibinfo{person}{Mingyang Yu}, \bibinfo{person}{Wei Zhan},
  \bibinfo{person}{Haohua Meng}, \bibinfo{person}{Chen Chen},
  \bibinfo{person}{Dai Hou}, {and} \bibinfo{person}{Junguo Hu}.}
  \bibinfo{year}{2023}\natexlab{}.
\newblock \showarticletitle{PoC Self-Generation Technology Based on
  Vulnerability Verification Program}. In \bibinfo{booktitle}{\emph{2023 {IEEE}
  4th Annual Flagship India Council International Subsections Conference
  ({INDISCON})}}. \bibinfo{publisher}{IEEE}, \bibinfo{address}{Mysore, India},
  \bibinfo{pages}{1--6}.
\newblock
\urldef\tempurl%
\url{https://doi.org/10.1109/INDISCON58499.2023.10270363}
\showDOI{\tempurl}


\bibitem[\protect\citeauthoryear{Madaan, Hermann, and Yazdanbakhsh}{Madaan
  et~al\mbox{.}}{2023}]%
        {madaan-etal-2023-makes}
\bibfield{author}{\bibinfo{person}{Aman Madaan}, \bibinfo{person}{Katherine
  Hermann}, {and} \bibinfo{person}{Amir Yazdanbakhsh}.}
  \bibinfo{year}{2023}\natexlab{}.
\newblock \showarticletitle{What Makes Chain-of-Thought Prompting Effective? A
  Counterfactual Study}. In \bibinfo{booktitle}{\emph{Findings of the
  Association for Computational Linguistics: EMNLP 2023}},
  \bibfield{editor}{\bibinfo{person}{Houda Bouamor}, \bibinfo{person}{Juan
  Pino}, {and} \bibinfo{person}{Kalika Bali}} (Eds.).
  \bibinfo{publisher}{Association for Computational Linguistics},
  \bibinfo{address}{Singapore}, \bibinfo{pages}{1448--1535}.
\newblock
\urldef\tempurl%
\url{https://doi.org/10.18653/v1/2023.findings-emnlp.101}
\showDOI{\tempurl}


\bibitem[\protect\citeauthoryear{Ni, Shen, Yang, Zhu, and Wang}{Ni
  et~al\mbox{.}}{2024}]%
        {ni2024megavul}
\bibfield{author}{\bibinfo{person}{Chao Ni}, \bibinfo{person}{Liyu Shen},
  \bibinfo{person}{Xiaohu Yang}, \bibinfo{person}{Yan Zhu}, {and}
  \bibinfo{person}{Shaohua Wang}.} \bibinfo{year}{2024}\natexlab{}.
\newblock \showarticletitle{MegaVul: A {C/C++} Vulnerability Dataset with
  Comprehensive Code Representations}. In \bibinfo{booktitle}{\emph{Proceedings
  of the 21st International Conference on Mining Software Repositories ({MSR}
  2024), Data and Tool Showcase}}. \bibinfo{publisher}{ACM}.
\newblock
\urldef\tempurl%
\url{https://doi.org/10.1145/3643991.3644886}
\showDOI{\tempurl}


\bibitem[\protect\citeauthoryear{{NIST}}{{NIST}}{2025}]%
        {site:NVD}
\bibfield{author}{\bibinfo{person}{{NIST}}.} \bibinfo{year}{2025}\natexlab{}.
\newblock \bibinfo{title}{National Vulnerability Database}.
\newblock \bibinfo{howpublished}{\url{https://nvd.nist.gov/}}.
\newblock
\newblock
\shownote{Accessed: 2025-07-29.}


\bibitem[\protect\citeauthoryear{{NVD - NIST}}{{NVD - NIST}}{2021}]%
        {cve-2021-25289}
\bibfield{author}{\bibinfo{person}{{NVD - NIST}}.}
  \bibinfo{year}{2021}\natexlab{}.
\newblock \bibinfo{booktitle}{\emph{CVE-2021-25289}}.
\newblock
\urldef\tempurl%
\url{https://nvd.nist.gov/vuln/detail/CVE-2021-25289}
\showURL{%
\tempurl}
\newblock
\shownote{Pillow TIFF decoding overflow; attributed to incomplete fix.}


\bibitem[\protect\citeauthoryear{{NVD - NIST}}{{NVD - NIST}}{2024}]%
        {cve-2024-27351}
\bibfield{author}{\bibinfo{person}{{NVD - NIST}}.}
  \bibinfo{year}{2024}\natexlab{}.
\newblock \bibinfo{booktitle}{\emph{CVE-2024-27351}}.
\newblock
\urldef\tempurl%
\url{https://nvd.nist.gov/vuln/detail/CVE-2024-27351}
\showURL{%
\tempurl}
\newblock
\shownote{Django ReDoS linked to incomplete fix of earlier CVEs.}


\bibitem[\protect\citeauthoryear{{OpenAI}}{{OpenAI}}{2023}]%
        {site:gpt-3.5}
\bibfield{author}{\bibinfo{person}{{OpenAI}}.} \bibinfo{year}{2023}\natexlab{}.
\newblock \bibinfo{title}{Function calling and other API updates}.
\newblock
  \bibinfo{howpublished}{\url{https://openai.com/index/function-calling-and-other-api-updates/}}.
\newblock
\newblock
\shownote{Accessed: 2025-07-29.}


\bibitem[\protect\citeauthoryear{{OpenAI}}{{OpenAI}}{2025}]%
        {site:o4-mini}
\bibfield{author}{\bibinfo{person}{{OpenAI}}.} \bibinfo{year}{2025}\natexlab{}.
\newblock \bibinfo{title}{Introducing OpenAI o3 and o4-mini}.
\newblock
  \bibinfo{howpublished}{\url{https://openai.com/index/introducing-o3-and-o4-mini/}}.
\newblock
\newblock
\shownote{Accessed: 2025-07-29.}


\bibitem[\protect\citeauthoryear{Orvalho and Kwiatkowska}{Orvalho and
  Kwiatkowska}{2025}]%
        {orvalho2025large}
\bibfield{author}{\bibinfo{person}{Pedro Orvalho} {and} \bibinfo{person}{Marta
  Kwiatkowska}.} \bibinfo{year}{2025}\natexlab{}.
\newblock \showarticletitle{Are Large Language Models Robust in Understanding
  Code Against Semantics-Preserving Mutations?}
\newblock \bibinfo{journal}{\emph{arXiv preprint arXiv:2505.10443}}
  (\bibinfo{year}{2025}).
\newblock


\bibitem[\protect\citeauthoryear{Simsek, Eghbali, and Pradel}{Simsek
  et~al\mbox{.}}{2025}]%
        {simsek2025pocgen}
\bibfield{author}{\bibinfo{person}{Deniz Simsek}, \bibinfo{person}{Aryaz
  Eghbali}, {and} \bibinfo{person}{Michael Pradel}.}
  \bibinfo{year}{2025}\natexlab{}.
\newblock \bibinfo{title}{PoCGen: Generating Proof-of-Concept Exploits for
  Vulnerabilities in {Npm} Packages}.
\newblock
\newblock
\showeprint[arxiv]{2506.04962}~[cs.CR]
\urldef\tempurl%
\url{https://arxiv.org/abs/2506.04962}
\showURL{%
\tempurl}


\bibitem[\protect\citeauthoryear{{VulnRepairEval}}{{VulnRepairEval}}{2025}]%
        {site:vulbench}
\bibfield{author}{\bibinfo{person}{{VulnRepairEval}}.}
  \bibinfo{year}{2025}\natexlab{}.
\newblock \bibinfo{title}{VulnRepairEval}.
\newblock \bibinfo{howpublished}{\url{\codelink}}.
\newblock
\newblock
\shownote{Accessed: 2025-07-29.}


\bibitem[\protect\citeauthoryear{Wang, Liu, and Xiao}{Wang
  et~al\mbox{.}}{2025}]%
        {wang2025cvebench}
\bibfield{author}{\bibinfo{person}{Peiran Wang}, \bibinfo{person}{Xiaogeng
  Liu}, {and} \bibinfo{person}{Chaowei Xiao}.} \bibinfo{year}{2025}\natexlab{}.
\newblock \showarticletitle{{CVE}-Bench: Benchmarking {LLM}-based Software
  Engineering Agent’s Ability to Repair Real-World {CVE} Vulnerabilities}. In
  \bibinfo{booktitle}{\emph{Proceedings of the 2025 Conference of the Nations
  of the Americas Chapter of the Association for Computational Linguistics:
  Human Language Technologies (Volume 1: Long Papers)}}.
  \bibinfo{publisher}{Association for Computational Linguistics},
  \bibinfo{address}{Albuquerque, USA}, \bibinfo{pages}{4207--4224}.
\newblock
\urldef\tempurl%
\url{https://aclanthology.org/2025.naacl-long.212/}
\showURL{%
\tempurl}


\bibitem[\protect\citeauthoryear{Wang, Li, Qian, Yang, Wang, Shang, Kumar, Tan,
  Ray, Bhatia, Nallapati, Ramanathan, Roth, and Xiang}{Wang
  et~al\mbox{.}}{2023}]%
        {wang-etal-2023-recode}
\bibfield{author}{\bibinfo{person}{Shiqi Wang}, \bibinfo{person}{Zheng Li},
  \bibinfo{person}{Haifeng Qian}, \bibinfo{person}{Chenghao Yang},
  \bibinfo{person}{Zijian Wang}, \bibinfo{person}{Mingyue Shang},
  \bibinfo{person}{Varun Kumar}, \bibinfo{person}{Samson Tan},
  \bibinfo{person}{Baishakhi Ray}, \bibinfo{person}{Parminder Bhatia},
  \bibinfo{person}{Ramesh Nallapati}, \bibinfo{person}{Murali~Krishna
  Ramanathan}, \bibinfo{person}{Dan Roth}, {and} \bibinfo{person}{Bing Xiang}.}
  \bibinfo{year}{2023}\natexlab{}.
\newblock \showarticletitle{{R}e{C}ode: Robustness Evaluation of Code
  Generation Models}. In \bibinfo{booktitle}{\emph{Proceedings of the 61st
  Annual Meeting of the Association for Computational Linguistics (Volume 1:
  Long Papers)}}, \bibfield{editor}{\bibinfo{person}{Anna Rogers},
  \bibinfo{person}{Jordan Boyd-Graber}, {and} \bibinfo{person}{Naoaki Okazaki}}
  (Eds.). \bibinfo{publisher}{Association for Computational Linguistics},
  \bibinfo{address}{Toronto, Canada}, \bibinfo{pages}{13818--13843}.
\newblock
\urldef\tempurl%
\url{https://doi.org/10.18653/v1/2023.acl-long.773}
\showDOI{\tempurl}


\bibitem[\protect\citeauthoryear{Wang, Hu, Gao, et~al\mbox{.}}{Wang
  et~al\mbox{.}}{2024}]%
        {wang2024reposvul}
\bibfield{author}{\bibinfo{person}{Xing Wang}, \bibinfo{person}{Ruiqi Hu},
  \bibinfo{person}{Chao Gao}, {et~al\mbox{.}}} \bibinfo{year}{2024}\natexlab{}.
\newblock \showarticletitle{ReposVul: A Repository-Level High-Quality
  Vulnerability Dataset}. In \bibinfo{booktitle}{\emph{Proceedings of the 2024
  {IEEE}/{ACM} 46th International Conference on Software Engineering: Companion
  Proceedings ({ICSE} Companion 2024)}}. \bibinfo{publisher}{IEEE},
  \bibinfo{pages}{472--483}.
\newblock
\urldef\tempurl%
\url{https://doi.org/10.1145/3639478.3647635}
\showDOI{\tempurl}


\bibitem[\protect\citeauthoryear{Wei, Wang, Schuurmans, Bosma, Ichter, Xia,
  Chi, Le, and Zhou}{Wei et~al\mbox{.}}{2022}]%
        {10.5555/3600270.3602070}
\bibfield{author}{\bibinfo{person}{Jason Wei}, \bibinfo{person}{Xuezhi Wang},
  \bibinfo{person}{Dale Schuurmans}, \bibinfo{person}{Maarten Bosma},
  \bibinfo{person}{Brian Ichter}, \bibinfo{person}{Fei Xia},
  \bibinfo{person}{Ed~H. Chi}, \bibinfo{person}{Quoc~V. Le}, {and}
  \bibinfo{person}{Denny Zhou}.} \bibinfo{year}{2022}\natexlab{}.
\newblock \showarticletitle{Chain-of-thought prompting elicits reasoning in
  large language models}. In \bibinfo{booktitle}{\emph{Proceedings of the 36th
  International Conference on Neural Information Processing Systems}} (New
  Orleans, LA, USA) \emph{(\bibinfo{series}{NIPS '22})}.
  \bibinfo{publisher}{Curran Associates Inc.}, \bibinfo{address}{Red Hook, NY,
  USA}, Article \bibinfo{articleno}{1800}, \bibinfo{numpages}{14}~pages.
\newblock
\showISBNx{9781713871088}


\bibitem[\protect\citeauthoryear{Wei, Duchenne, Copet, et~al\mbox{.}}{Wei
  et~al\mbox{.}}{2025}]%
        {wei2025swerl}
\bibfield{author}{\bibinfo{person}{Yizheng Wei}, \bibinfo{person}{Olivier
  Duchenne}, \bibinfo{person}{Jade Copet}, {et~al\mbox{.}}}
  \bibinfo{year}{2025}\natexlab{}.
\newblock \bibinfo{title}{{SWE}-RL: Advancing {LLM} Reasoning via Reinforcement
  Learning on Open Software Evolution}.
\newblock
\newblock
\showeprint[arxiv]{2502.18449}~[cs.LG]
\urldef\tempurl%
\url{https://arxiv.org/abs/2502.18449}
\showURL{%
\tempurl}


\bibitem[\protect\citeauthoryear{Widyasari, Sim, Lok, Qi, Phan, Tay, Tan, Wee,
  Tan, Yieh, Goh, Thung, Kang, Hoang, Lo, and Ouh}{Widyasari
  et~al\mbox{.}}{2020}]%
        {widyasari2020bugsinpy}
\bibfield{author}{\bibinfo{person}{Ratnadira Widyasari},
  \bibinfo{person}{Sheng~Qin Sim}, \bibinfo{person}{Camellia Lok},
  \bibinfo{person}{Haodi Qi}, \bibinfo{person}{Jack Phan},
  \bibinfo{person}{Qijin Tay}, \bibinfo{person}{Constance Tan},
  \bibinfo{person}{Fiona Wee}, \bibinfo{person}{Jodie~Ethelda Tan},
  \bibinfo{person}{Yuheng Yieh}, \bibinfo{person}{Brian Goh},
  \bibinfo{person}{Ferdian Thung}, \bibinfo{person}{Hong~Jin Kang},
  \bibinfo{person}{Thong Hoang}, \bibinfo{person}{David Lo}, {and}
  \bibinfo{person}{Eng~Lieh Ouh}.} \bibinfo{year}{2020}\natexlab{}.
\newblock \showarticletitle{BugsInPy: A Database of Existing Bugs in Python
  Programs to Enable Controlled Testing and Debugging Studies}. In
  \bibinfo{booktitle}{\emph{Proceedings of the 28th {ACM} Joint Meeting on
  European Software Engineering Conference and Symposium on the Foundations of
  Software Engineering ({ESEC}/{FSE} 2020)}}. \bibinfo{publisher}{ACM},
  \bibinfo{address}{New York, NY, USA}, \bibinfo{pages}{1556--1560}.
\newblock
\urldef\tempurl%
\url{https://doi.org/10.1145/3368089.3417943}
\showDOI{\tempurl}


\bibitem[\protect\citeauthoryear{Yang, Li, Yang, Zhang, Hui, Zheng, Yu, Gao,
  Huang, Lv, et~al\mbox{.}}{Yang et~al\mbox{.}}{2025}]%
        {yang2025qwen3}
\bibfield{author}{\bibinfo{person}{An Yang}, \bibinfo{person}{Anfeng Li},
  \bibinfo{person}{Baosong Yang}, \bibinfo{person}{Beichen Zhang},
  \bibinfo{person}{Binyuan Hui}, \bibinfo{person}{Bo Zheng},
  \bibinfo{person}{Bowen Yu}, \bibinfo{person}{Chang Gao},
  \bibinfo{person}{Chengen Huang}, \bibinfo{person}{Chenxu Lv},
  {et~al\mbox{.}}} \bibinfo{year}{2025}\natexlab{}.
\newblock \showarticletitle{Qwen3 technical report}.
\newblock \bibinfo{journal}{\emph{arXiv preprint arXiv:2505.09388}}
  (\bibinfo{year}{2025}).
\newblock


\bibitem[\protect\citeauthoryear{Yang, Jimenez, Wettig, Lieret, Yao,
  Narasimhan, and Press}{Yang et~al\mbox{.}}{2024}]%
        {yang2024sweagent}
\bibfield{author}{\bibinfo{person}{John Yang}, \bibinfo{person}{Carlos~E
  Jimenez}, \bibinfo{person}{Alexander Wettig}, \bibinfo{person}{Kilian
  Lieret}, \bibinfo{person}{Shunyu Yao}, \bibinfo{person}{Karthik~R
  Narasimhan}, {and} \bibinfo{person}{Ofir Press}.}
  \bibinfo{year}{2024}\natexlab{}.
\newblock \showarticletitle{{SWE}-agent: Agent-Computer Interfaces Enable
  Automated Software Engineering}. In \bibinfo{booktitle}{\emph{The
  Thirty-eighth Annual Conference on Neural Information Processing Systems}}.
\newblock
\urldef\tempurl%
\url{https://arxiv.org/abs/2405.15793}
\showURL{%
\tempurl}


\bibitem[\protect\citeauthoryear{Zhuo, Vu, Chim, et~al\mbox{.}}{Zhuo
  et~al\mbox{.}}{2024}]%
        {zhuo2024bigcodebench}
\bibfield{author}{\bibinfo{person}{Tianyi Zhuo}, \bibinfo{person}{Minh~Chien
  Vu}, \bibinfo{person}{Jingkai Chim}, {et~al\mbox{.}}}
  \bibinfo{year}{2024}\natexlab{}.
\newblock \bibinfo{title}{BigCodeBench: Benchmarking Code Generation with
  Diverse Function Calls and Complex Instructions}.
\newblock
\newblock
\showeprint[arxiv]{2406.15877}~[cs.LG]
\urldef\tempurl%
\url{https://arxiv.org/abs/2406.15877}
\showURL{%
\tempurl}


\end{thebibliography}
	
\end{document}